\documentclass{aastex62}

\usepackage{graphicx,blindtext,rotating}
\usepackage{ amssymb }

\newcommand{\gtt}{\hbox{G23.11$+$0.18}}
\newcommand{\hje}{\hbox{HESS\,J1832$-$085}}
\newcommand{\RXJ}{\hbox{RX\,J1713.7$-$3946}}
\newcommand{\Jsto}{\hbox{HESS\,J1731$-$347}}

\newcommand{\HII}{{H\,{\sc ii}}\xspace}

\newcommand{\distance}{\hbox{4.6$\pm$0.8\,kpc}}

\newcommand{\perbeam}{\,beam\ensuremath{^{-1}}}

\newcommand{\newtext}[1]{\textcolor{black}      {#1}}

\newcommand{\UNSWcan}{School of Science, The University of New South Wales, Australian Defence Force Academy, Canberra 2610, Australia}
\newcommand{\WSU}{Western Sydney University, Locked Bag 1797, Penrith, NSW 2751, Australia}
\newcommand{\ICRAR}{International Centre for Radio Astronomy Research, Curtin University, Bentley, WA 6102, Australia}
\newcommand{\UoA}{School of Physical Sciences, The University of Adelaide, Adelaide 5005, Australia}
\newcommand{\MPIG}{Max-Planck-Institut f\"{u}r extraterrestrische Physik, Giessenbachstra\ss e, 85748 Garching, Germany}
\newcommand{\UNSW}{School of Physics, The University of New South Wales, Sydney 2052, Australia}
\newcommand{\Arm}{Armagh Observatory and Planetarium, College Hill, Armagh BT61 9DG, UK}
\newcommand{\Tueb}{Institut f\"{u}r Astronomie und Astrophysik, Universit\"{a}t T\"{u}bingen, 72076 T\"{u}bingen, Germany}
\newcommand{\Nag}{Department of Physics, Nagoya University, Furo-cho, Chikusa-ku, Nagoya 464-8601, Japan}
\newcommand{\Bam}{Dr. Karl Remeis Observatory and ECAP, Universit\"{a}t Erlangen-N\"{u}rnberg, Sternwartstr. 7, 96049 Bamberg, Germany}
\newcommand{\NAOB}{National Astronomical Observatories, CAS, Beijing 100012, China}
\newcommand{\UCAB}{University of Chinese Academy of Sciences, Beijing 100049, China}
\newcommand{\Calgary}{Department of Physics \& Astronomy, University of Calgary, Calgary, Alberta T2N 1N4, Canada}
\newcommand{\CAASTRO}{ARC Centre of Excellence for All-sky Astrophysics (CAASTRO)}

\graphicspath{{./}{figures/}}

\received{July 3, 2019}
\revised{August 18, 2019}
\accepted{August 23, 2019}
\submitjournal{ApJ}

\shorttitle{The Galactic SNR \gtt}

\shortauthors{Maxted et al.}

\begin{document}

\title{{A Supernova Remnant Counterpart for HESS\,J1832-085}}

\correspondingauthor{Nigel I. Maxted}
\email{n.maxted.astro@gmail.com}

\author[0000-0003-2762-8378]{Nigel I. Maxted}
\affil{\UNSWcan}

\author[0000-0002-4990-9288]{M.~D.~Filipovi\'c}
\affil{\WSU}

\author[0000-0002-5119-4808]{N.~Hurley-Walker}
\affil{\ICRAR}

\author[0000-0002-4362-9664]{I.~Boji\v{c}i\'c}
\affil{\WSU} 

\author[0000-0002-9516-1581]{G.~P.~Rowell}
\affil{\UoA}

\author[0000-0002-0107-5237]{F.~Haberl}
\affil{\MPIG} 

\author[0000-0002-4794-6835]{A.~J.~Ruiter}
\affil{\UNSWcan}

\author[0000-0002-5044-2988]{I.~R.~Seitenzahl}
\affil{\UNSWcan}

\author[0000-0002-2618-5627]{F.~Panther}
\affil{\UNSWcan}

\author{G.~F.~Wong}
\affil{\WSU}
\affil{\UNSW}

\author[0000-0002-6475-0797]{C.~Braiding}
\affil{\UNSW}

\author[0000-0001-7289-1998]{M.~Burton}
\affil{\Arm}

\author{G.~P\"{u}hlhofer}
\affil{\Tueb}

\author[0000-0003-2062-5692]{H.~Sano}
\affil{\Nag}

\author{Y.~Fukui}
\affil{\Nag}

\author[0000-0001-5302-1866]{M.~Sasaki}
\affil{\Bam}

\author{W.~Tian}
\affil{\NAOB}
\affil{\UCAB}

\author{H.~Su}
\affil{\NAOB}

\author{X.~Cui}
\affil{\NAOB}

\author[0000-0002-4814-958X]{D.~Leahy}
\affil{\Calgary}





\author[0000-0002-4203-2946]{P.~J.~Hancock}
\affil{\ICRAR}
\affil{\CAASTRO}

\begin{abstract}
We examine the new Galactic supernova remnant (SNR) candidate, \gtt, as seen by the Murchison Widefield Array (MWA) radio telescope. We describe the morphology of the candidate and find a spectral index of $-$0.63$\pm$0.05 in the 70-170\,MHz domain. A \newtext{coincident} TeV gamma-ray detection in High-Energy Stereoscopic System (HESS) data {supports the SNR nature of \gtt\ and suggests that \gtt\ is} accelerating particles beyond TeV energies, thus making this object a promising new cosmic ray hadron source candidate. The remnant cannot be seen in current optical, infrared and X-ray data-sets. We do find, however, a dip in CO-traced molecular gas at a line-of-sight velocity of $\sim$85\,km\,s$^{-1}$, suggesting the existence of a \gtt\ progenitor wind-blown bubble. 
\newtext{Furthermore, the discovery of molecular gas clumps at a neighbouring velocity towards \hje\ adheres to the notion that a hadronic gamma-ray production mechanism is plausible towards the north of the remnant. }
Based on these morphological arguments, we propose an interstellar medium association for \gtt\ at a kinematic distance of \distance . 
\end{abstract}




\section{INTRODUCTION }
\label{sec:intro}
Supernova remnants (SNRs) are key candidates in the search for the Galaxy's sources of cosmic ray hadrons up to energies of several PeV.  
Investigations of cosmic-ray \newtext{(CR)} origin are partly enabled by gamma-ray instruments, which can trace distant populations of high-energy particles through interaction byproducts. Similarly, non-thermal radio continuum emission traces populations of high energy electrons as they gain and then lose energy in the shells of SNRs. It follows that an inherent connection between radio emission and gamma-rays exists in the non-thermal spectra of SNRs, thus together, these two energy-windows can deliver constraints on particle acceleration {and are complimentary in identifying SNRs \citep[e.g.][]{Abramowski:2017newshells}. We place the new supernova remnant \gtt\ (the topic of this paper) and coincident TeV gamma-ray source, \hje , in this category.}

{In several middle-aged \newtext{($\sim$10$^4$\,yr)} SNRs, the Fermi gamma-ray space telescope successfully measured a distinctive high-energy spectral feature that is related to \newtext{CR} proton interactions with gas - the so-called `pion-bump' \citep{Ackermann:2013} from proton-proton {(p-p)} interactions that create neutral pions, which subsequently decay into gamma-ray photons. }
The process naturally connects interstellar medium (ISM) gas and gamma-rays, and has allowed several other middle-aged SNRs to be identified as past CR accelerators through the observed overlap between TeV gamma-rays and molecular gas, which is impacted by high-energy protons diffusing away from their acceleration sites \citep[e.g.][]{Aharonian:w28,Albert:2007_IC443}. 
{P-p interactions of CR protons diffusing} away from a local source has also been invoked to explain an excess of gamma-ray emission from the Galactic Centre, given an energy-dependent distribution measured by {High-Energy Stereoscopic System (HESS)} \citep{Abramowski:2016nature}. 

Gamma-ray emission can also be created via the inverse-Compton mechanism, whereby high-energy electrons up-scatter low-energy photons to gamma-ray energies. The gamma-ray mechanism of some young shell-type gamma-ray-bright SNRs is presently the subject of vigorous debate. Attempts to replicate global gamma-ray spectra may favour electrons as the primary dominant particle triggering gamma-ray emission -- a so-called `leptonic' model \citep[e.g. see][]{Acero:2015}, while conversely, good correspondences between ISM gas density and gamma-ray flux is suggestive of CR proton interactions -- a so-called `hadronic' model in the very same young objects \citep[][for \RXJ, \Jsto\ and Vela\,Jr, respectively]{Fukui:2012,Fukuda:2014,Fukui:2017}\footnote{However, evidence for a dominant hadronic component for \Jsto\ only exists for the 5.2-6\,kpc distance solution, as opposed to the 3.2\,kpc solution favoured by the \citet{Cui:2016} model and explicitly derived in \citet{Maxted:2018_HESSJ1731}}. 
{In many cases, the absence of direct shock tracers \citep[such as SiO emission for example,][]{Nicholas:2012} means that proposed ISM associations primarily rely on matching high energy features to gas structure. }

Extensive modeling has shown that a gamma-ray spectral shape similar to leptonic gamma-ray emission
can plausibly be created by hadronic processes in the presence of an inhomogenous and clumpy circumstellar medium \citep[e.g.][]{Zirakashvili:2010,Inoue:2012,Fukui:2012,Celli:2018arXiv,Inoue:2019}{, hence resolving the two seemingly-conflicting lines of evidence. 
In contrast with the aforementioned escaping CR scenario, this would be hadronic gamma-ray emission from within the SNR shell boundary rather than outside the shell.}
{Another way to resolve the leptonic high energy spectral shape of young shell-type gamma-ray SNRs with the observed gas/gamma-ray correlation characteristic of hadronic gamma-ray emission may be to model electron injection as a function of ISM density \citep[e.g. as suggested in][]{Sushch:2018}. The nature of gamma-ray emission (lepton or hadron-dominated) from young SNRs {is a matter} of ongoing debate. }

Regardless of the dominant gamma-ray emission mechanism, acceleration of electrons to TeV energies is occurring in young SNRs, as evidenced by synchrotron X-ray emission from young SNR shocks. The non-thermal spectrum largely follows the energy spectrum of the local electron population, often extending down to radio frequencies. In this way, the radio, X-ray and gamma-ray skies are inherently linked, and a characterisation of the radio spectrum can provide constraints on electron acceleration within SNRs. The processes of proton acceleration are expected to be quite similar, {so such constraints are also very relevant to investigations of CR origin, and either adjacent or embedded gas can help us form an understanding of the nature of associated gamma-ray emission (as is the case for the aforementioned examples of W28 and \RXJ , respectively).} 

{Finally, we note that} the discovery of new Galactic SNRs generally, by any means, helps minimise the discrepancy observed between star formation rates and the population of known SNRs \citep[e.g. see][]{Brogan:2006,2007MNRAS.374.1441S,2007MNRAS.381..377S,2017ifs..confE.101E}.

In this paper, we investigate the new SNR \gtt\  in the context of the search for Galactic CR hadron sources that are not necessarily expected to have clear shock-interaction signatures in an inhomogeneous interstellar medium. In Section\,\ref{sec:w41}, we summarise the current knowledge of the \gtt\ region. In Section\,\ref{sec:Obs}, we outline the multi-wavelength (MWL) data analysed in this study, and in Section\,\ref{sec:ResDisc} we present new images and spectra of \gtt\ before discussing our search for MWL counterparts. We conclude by proposing a connection between \gtt , \hje\ and the interstellar medium.

\begin{figure}
\centering
\includegraphics[trim={0 0 3.0cm 0},clip,width=0.58\textwidth]{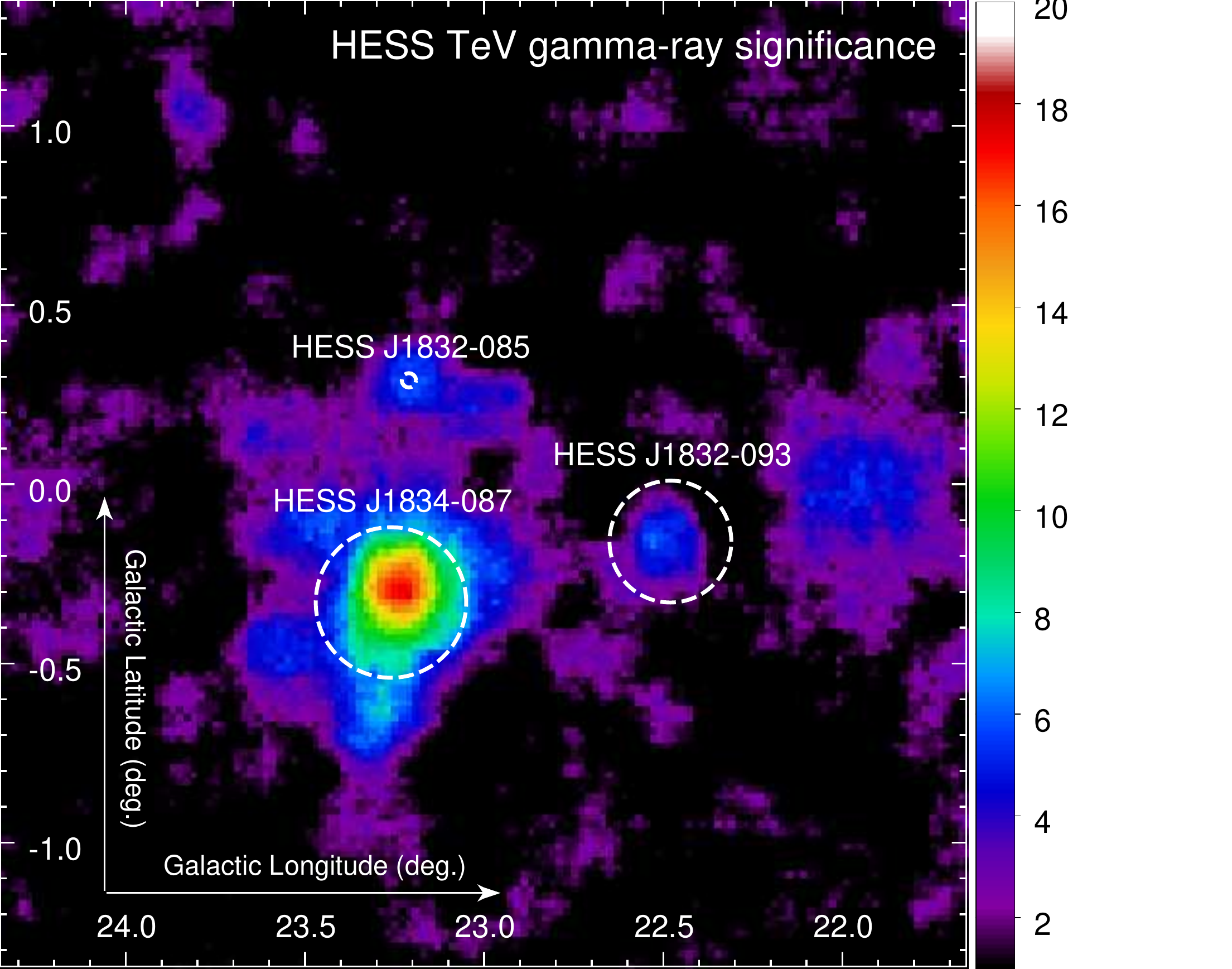}
\includegraphics[width=0.58\textwidth]{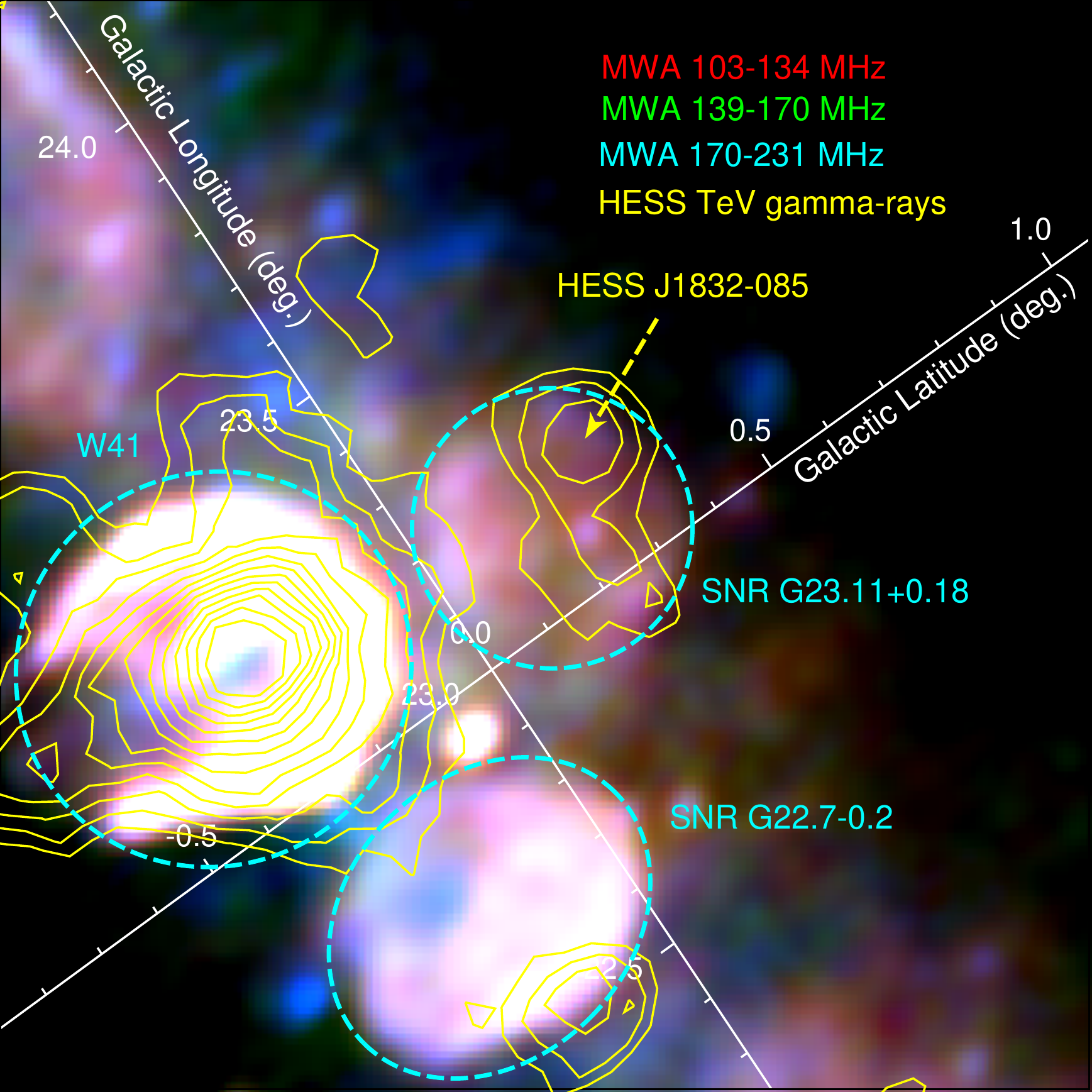}\\
\caption{\textbf{Top:} HESS telescope Galactic Plane Survey image of TeV gamma-ray emission significance towards the HESS\,J1834$-$087 region \citep[see][for details]{Abdalla:2018}. Previously established/known SNRs are marked with dashed circles. \label{fig:HESSsig} \textbf{Bottom:} {3-colour image of 3 sub-bands of Murchison Widefield Array radio continuum data (red:103-134\,MHz, green:139-170\,MHz, blue:170-231\,MHz) towards the region containing the new SNR candidate \gtt. \gtt, composite PWN/SNR W41 and SNR G22.7-0.2 are indicated by dashed cyan ellipses. HESS TeV gamma-ray emission significance contours \citep{Abdalla:2018}, starting at a 3$\sigma$ level, are overlaid.}
 \label{fig:MWAallbands}}
\end{figure}

\subsection{The W41 region and \hje }\label{sec:w41}
Figure\,\ref{fig:HESSsig} is an image of gamma-ray emission towards HESS\,J1834$-$087 and the surrounding region \citep{Abdalla:2018}. Bright gamma-ray emission from the composite SNR/pulsar-wind-nebula system, W41, is known as HESS\,J1834$-$087 and dominates the image. Several weaker gamma-ray sources exist nearby. To the centre-north is the unidentified {`point-like'} gamma-ray source, \hje . The HESS collaboration investigated a possible pulsar (PSR\,J1832$-$0836/MSP\,J1832$-$0836
) origin for this source. {Although the spin-down power is not considered sufficient to power such luminous gamma-ray emission, for millisecond pulsars that may have been `spun-up' by accretion \citep[e.g.][]{0004-637X-864-1-23} the precise relationship between spin-down power and luminosity is unclear. }
{We revisit this scenario in Section\,\ref{sec:Distance}, after we examine \hje\ in light of a newly-discovered SNR (\gtt), as seen in Murchison Widefield Array (MWA) data in Section\,\ref{sec:Rad}.}

To the west of HESS\,J1834$-$087 is HESS\,J1832$-$093. Recent observations, strongly indicate a non-accreting binary origin for this gamma-ray emission \citep{Eger:2016,Mori:2017}. This is contrasted with initial investigations which suggested a possible association with the SNR\,G22.7$-$0.2 \citep{Abramowski:2015_HESSJ1832-085}. \citet{Su:2014} previously identified a possible molecular cloud counterpart at 75-79\,km~s$^{-1}$, which was at the time suggested to be a target of CR hadrons accelerated in the shock of SNR\,G22.7$-$0.2. However, this SNR-cloud interaction origin for HESS\,J1834$-$087 is now disfavoured \citep{Abramowski:2015_HESSJ1832-085}. \citet{Su:2014} also noted that the cloud\footnote{The authors suggested a 4.4$\pm$0.4\,kpc distance and Scutum-Crux arm association for this cloud, but we suggest that a Norma-arm association beyond 70\,km~s$^{-1}$ is more likely.} is a possible candidate for association with the W41 SNR and the \HII\ region G022.760$-$0.485. Indeed, gas at this velocity is also considered in our examination of potential associations for the recently discovered SNR candidate \gtt\ in Section\,\ref{sec:Gas}.

\section{Observations}\label{sec:Obs}
In this paper we primarily utilise radio continuum data from the Murchison Widefield Array (MWA), TeV gamma-ray data from the {High Energy Stereoscopic System (HESS)} Galactic Plane Survey \citep{Abdalla:2018}, molecular gas data from the FUGIN CO(1-0) survey \citep{Umemoto:2017_FUGIN1} {and X-ray data from the X-ray Multi-Mirror Mission (a.k.a \textit{XMM-Newton}).}
Several instances of SNR candidate overlap with gamma-ray emission were present in a preliminary investigation carried out using MWA and HESS data. In this study we solely focus on one compelling source, the SNR candidate \gtt .

\subsection{MWA radio continuum}
The Murchison Widefield Array (MWA) is a collection of 2048 dual-polarisation dipole antennas situated in the Shire of Murchison in Western Australia, approximately 300\,km from the small coastal town of Geraldton. The MWA is the low-frequency pre-cursor to the Square Kilometer Array (SKA) and is comprised of a central core of dipoles within a $\sim$1.5\,km region with extensions out to 3\,km to cover longer baselines. The MWA has no moving parts such that the array is `steered' electronically within 25$^{\circ}$ of sky \citep[see][for details]{Tingay:2013}.

Data from the GaLactic and Extragalactic All-sky Murchison Widefield Array (GLEAM) survey \citep{Wayth:2015} was processed to produce a data release covering covering $345^\circ < l < 60^\circ$, $180^\circ < l < 240^\circ$, $|b|\leq 10^\circ$ \citet{Hurley-Walker_DR:2019}, which has an effective integration time of about 10~minutes per pixel. \citet{Hurley-Walker_cand:2019} examined this data for known Galactic SNR candidates, and \citet{Hurley-Walker_new:2019} searched it for new SNRs, finding 26. The GLEAM data-set is sensitive to radio continuum emission at frequencies between 72 and 231\,MHz and is made up of 20 bands with a channel bandwidth of 7.68\,MHz. For faint objects, it is useful to combine the data into 4~bins: 72--103, 103--134, 139--170 and 170--231\,MHz, with corresponding image Point Spread Functions (PSFs) of 5.2$^{\prime}$, 3.9$^{\prime}$, 2.9$^{\prime}$ and 2.4$^{\prime}$.

{New GLEAM MWA data is complemented by archival data from the Bonn 11-cm (2.695\,GHz) Survey with the Effelsberg Telescope \citep{Reich:1984_Effels}. This is a single-dish survey covering $357\fdg4 \leq l \leq 76^\circ$, $b\leq |1\fdg5|$ with $4\farcm3$ resolution and 50\,mK (20\,mJy\perbeam) sensitivity\footnote{ \href{http://www3.mpifr-bonn.mpg.de/survey.html}{http://www3.mpifr-bonn.mpg.de/survey.html}}. }

\subsection{HESS Gamma-rays}
The HESS telescope is a system of 5 ground-based Cherenkov light detectors operating in the Khomas Highland of Namibia at an altitude of 1800\,m. The array consists of four 12\,m diameter telescopes, operating from 2003, and one 28\,m diameter telescope, operating since 2012. The five detectors observe Cherenkov light from showers of high energy particles produced by $\sim$0.1-100\,TeV-energy gamma-ray photons interacting in the air above the dark Khomas Highland site. 

The HESS Galactic Plane Survey utilised in this work is a compilation of 14 years of observations with the four 12\,m-diameter telescopes. It has varying energy threshold and spatial resolution across the Milky Way, while the telescope is generally sensitive to photon energies larger than $>$200\,GeV with a photon energy-reconstruction accuracy of $\sim$15\%. The photon arrival direction reconstruction is accurate to within $\sim$0.08$^{\circ}$ \citep{Abdalla:2018}, which is the approximate angular resolution of resultant maps. 

\subsection{FUGIN CO(1-0)}
FUGIN is an acronym for the FOREST Unbiased Galactic plane Imaging survey with the Nobeyama 45\,m telescope, where FOREST is itself an acronym for the FOur-beam REceiver System on the 45-m Telescope \citep{Minamidani:2016}. The Nobeyama telescope is located at an elevation of 1350\,m in Minamimaki, near Nagano, Japan. 

The Nobeyama telescope has a beam {full-width half-maximum} of 14$^{\prime\prime}$ at 115\,GHz, which results in a post-reduction CO(1-0) angular resolution of 20$^{\prime\prime}$ and a $^{13}$CO/C$^{18}$O(1-0) angular resolution of 21$^{\prime\prime}$ for FUGIN survey data \citep{Umemoto:2017_FUGIN1}. The FUGIN CO(1-0) velocity resolution and beam efficiency are 1.3\,km~s$^{-1}$ and 0.43$\pm$0.02, respectively. 

As part of this investigation, FUGIN CO data is processed using Miriad software \citep{Sault:1995}.

We note that a search for gas counterparts for Galactic SNRs detected in the GLEAM survey will be viable using the southern hemisphere Mopra Galactic Plane CO survey \citep{Burton:2013,Braiding:2015,Braiding:2018} between longitudes of $-$110$^{\circ}$ and 11$^{\circ}$. In the particular case of \gtt , the candidate falls outside of the Mopra survey coverage. Generally, we anticipate the Mopra and FUGIN surveys, in the southern and northern hemispheres, respectively, to be complementary in future studies of SNRs in the Galactic Plane.

\subsection{{\it XMM-Newton} X-rays}\label{ssec:Xrayobs}
We also examined {\it XMM-Newton} 0.2-4.5\,keV X-ray emission data for signatures of the proposed SNR candidate. Approximately {30\,ksec} of {\it XMM-Newton} exposure were obtained in October 2007 towards the bright source AX~J1832.3-0840. The three {Eurpoean Photon Imaging Camera (EPIC)} instruments \citep{2001A&A...365L..27T,2001A&A...365L..18S} are operated simultaneously and cover a field of view of about 30$^{\prime}$ in diameter, encompassing a large fraction of \gtt. 

AX~J1832.3-0840 is most likely a magnetic cataclysmic variable (intermediate polar) or a high-mass X-ray binary. It shows X-ray pulsations with a period of $\sim$1550~s discovered in the Advanced Satellite for Cosmology and Astrophysics ({\it ASCA}) data \citep[][also see \citealt{Gaia:2018}]{2010MNRAS.402.2388K}.

We created {\it XMM-Newton} images following \citet{2013A&A...558A...3S}. After removal of two short background flares, the net exposure times for the three EPIC instruments\footnote{Two Metal-Oxide-Silicon detectors and a fully depleted PN CCD} were 26.0~ks, 28.6~ks and 28.8~ks for pn, MOS1 and MOS2, respectively.  

The detection of a potential diffuse X-ray emission signal from SNR candidate \gtt\ is hampered by high foreground absorption, which attenuates soft \mbox{X-rays}, and dust scattering halos around bright sources, particularly towards this dense and active part of the Galactic Plane. \newtext{Assuming a visual extinction to reddening ratio of 3.1 \citep{Schlafly:2011} and $\sim$2.1-2.2$\times$10$^{21}$\,cm$^{-2}$ of foreground H column density per magnitude of visual extinction \citep{Guver:2009,Zhu:2017}, the measured optical extinction\footnote{ http://irsa.ipac.caltech.edu/applications/DUST/} of 35-40\,$A_{\nu}$ towards \hje\ implies a possible foreground absorption column density of up to 7-9$\times$10$^{22}$\,cm$^{-2}$. For such a column density, a typical SNR with thermal temperatures of 0.25-0.6\,keV would have $>$99\% of $\lesssim$5\,keV X-ray emission absorbed by photoelectric absorption, assuming solar elemental abundances in foreground gas. We note that some proportion of visual extinction may be attributable to gas associated with \gtt , as discussed in Section\,\ref{sec:Infrared}. }
The results of our X-ray investigation are discussed in Section\,\ref{sec:Xray}.

\section{Results and Discussion} \label{sec:ResDisc}

The SNR candidate, \gtt, is indicated in Figure\,\ref{fig:MWAallbands}, which shows three frequency bands of MWA data towards the Galactic plane centred on the SNR. In the following subsection, we summarise the radio continuum characteristics of \gtt, before a multi-wavelength investigation is outlined in subsequent subsections. We particularly focus on the origin of the gamma-ray source \hje, in the light of the discovery of \gtt. {Finally, we propose and discuss the nature and distance of the new SNR candidate. }



\subsection{The \gtt\ Radio Continuum}
 \label{sec:Rad}
\begin{figure*}
\includegraphics[width=0.49\textwidth]{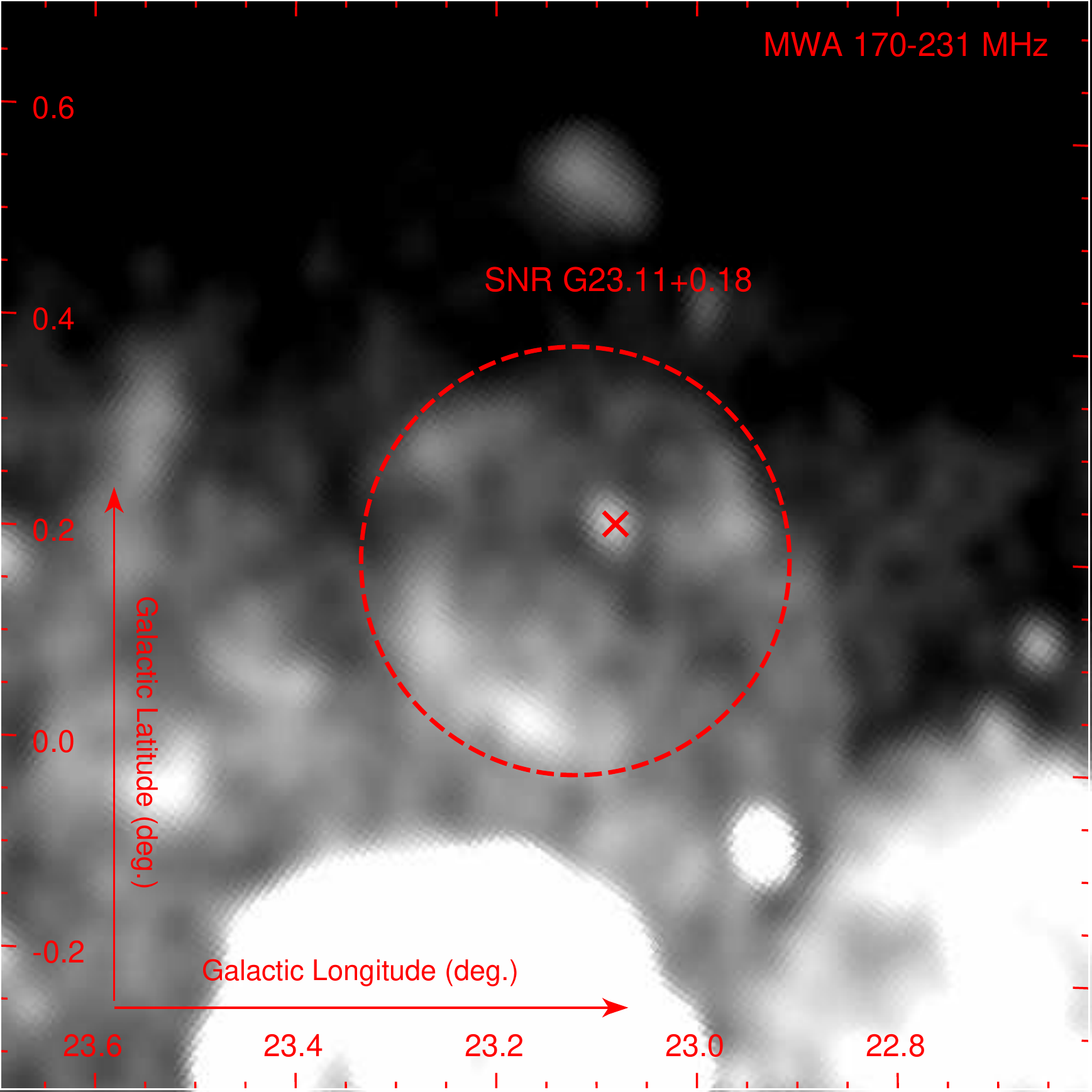}
\includegraphics[width=0.49\textwidth]{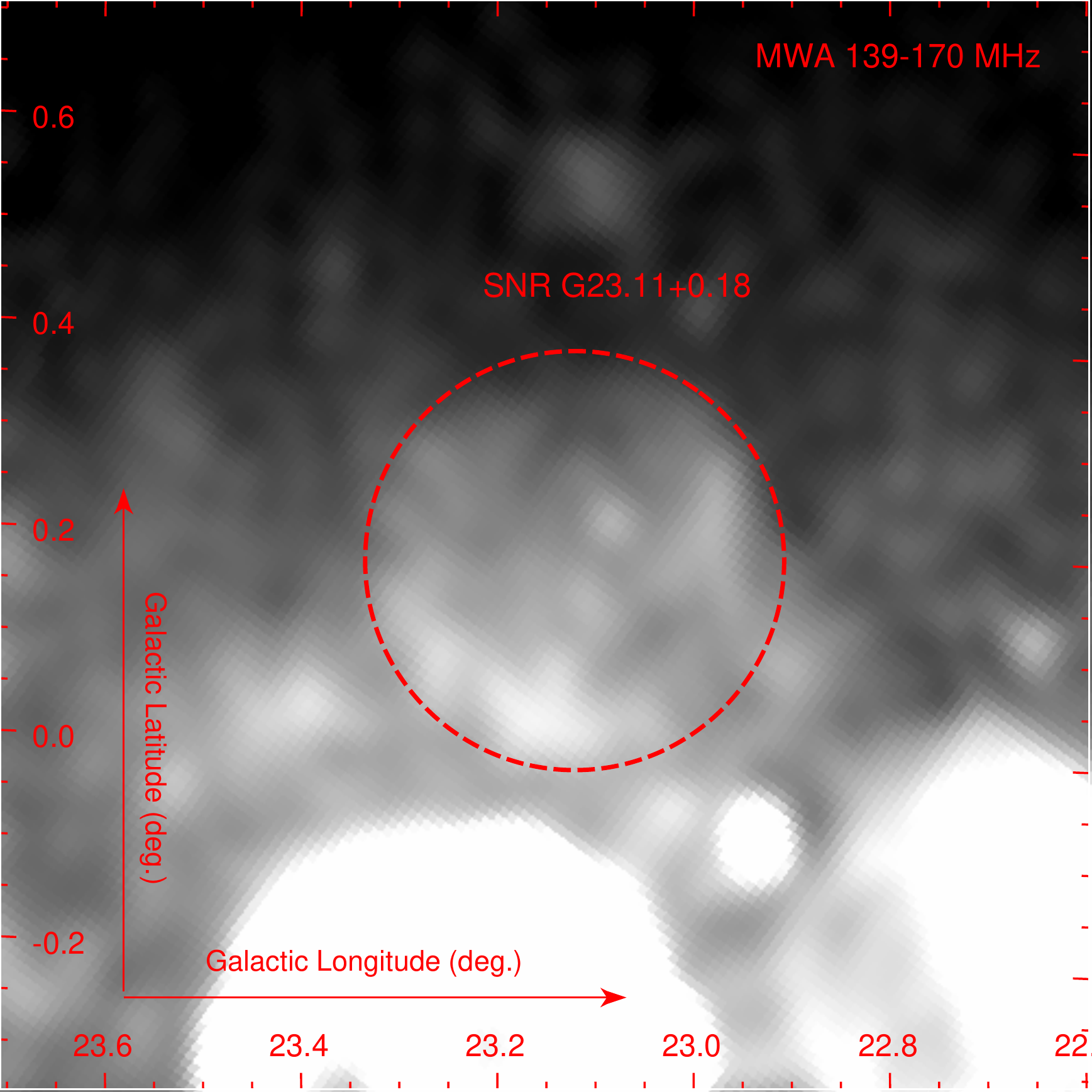}\\
\includegraphics[width=0.49\textwidth]{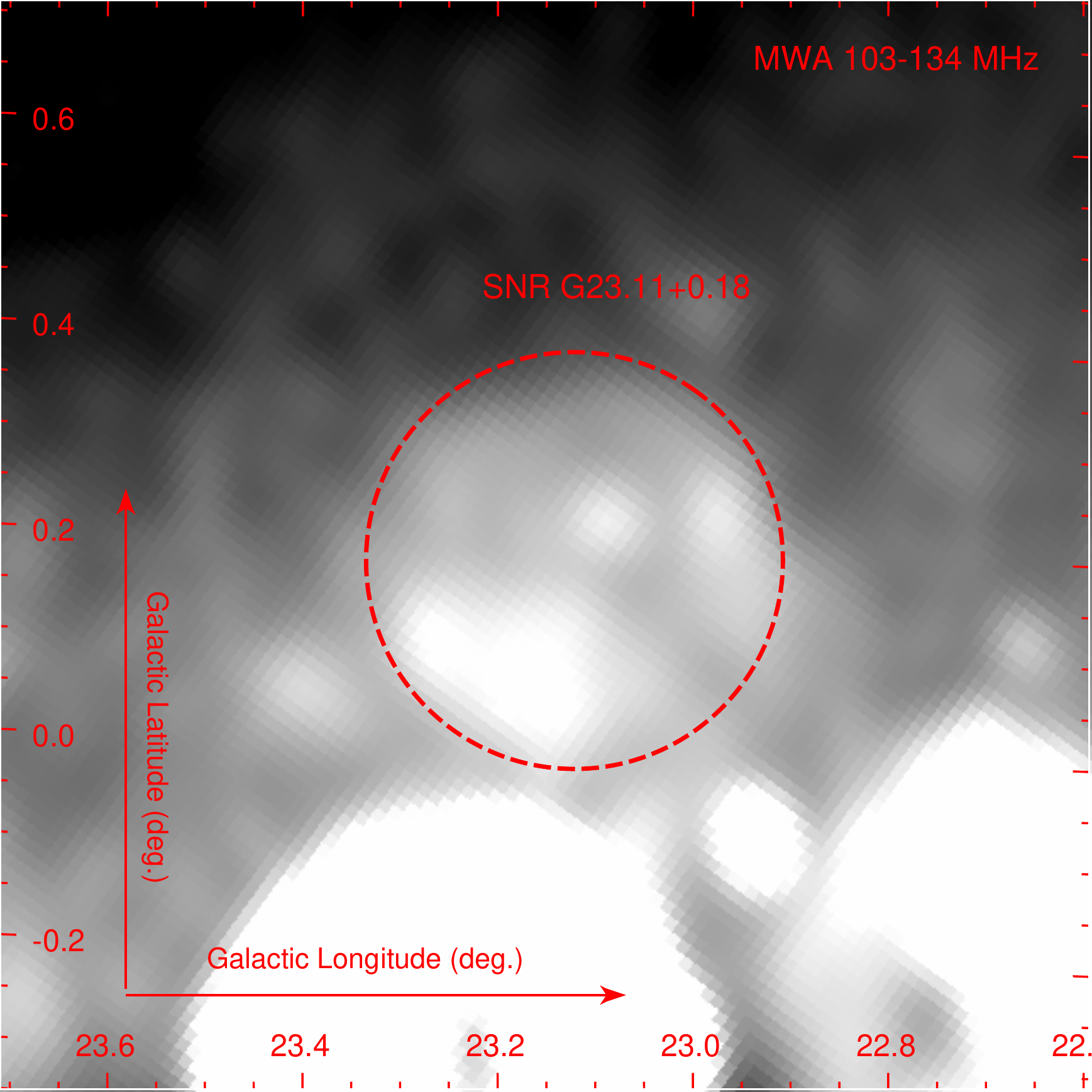}
\includegraphics[width=0.49\textwidth]{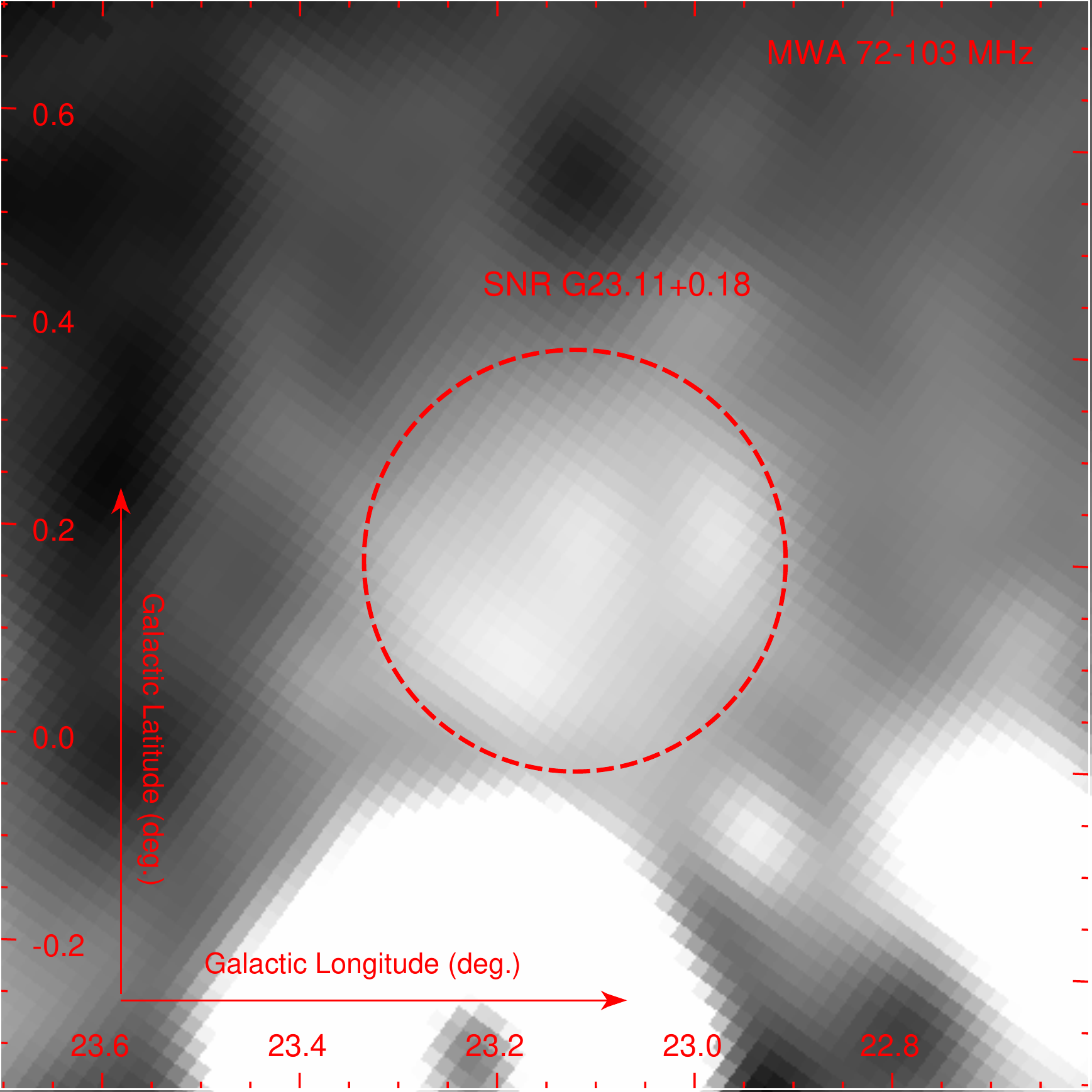}
\caption{Murchison Widefield Array radio continuum images of the new SNR candidate in 4 wavebands - 170--231, 139--170, 103--134 and 72--103\,MHz. A red dashed circle indicates the boundary of SNR candidate 23.1$+$0.2. In the top left image, the position of the radio continuum point source G23.08+0.22 is indicated by a cross.
 \label{fig:MWAindividual}}
\end{figure*}

Images of four frequency bands of MWA radio continuum data are displayed in Figure\,\ref{fig:MWAindividual}. The proposed extent of non-thermal emission from the shell of the new SNR candidate, previously designated SNR\,\gtt\ in \citet{Anderson:2017_THORsnrs} and independently discovered in our study, is indicated by a circle at centroid ($l$,$b$) of (23.12$^{\circ}$, 0.19$^{\circ}$ (RA (J2000), Dec (J2000): $\sim$ 18$^h$32$^m$29.7$^s$,--8$^{\circ}$39$^{\prime}$12.9$^{\prime\prime}$) and radius of 750$^{\prime\prime}$ in Figure\,\ref{fig:MWAallbands} and \ref{fig:MWAindividual}. We estimate the SNR radius {(with conservative uncertainty)} to be 700$\pm$50$^{\prime\prime}$. 

The apparent SNR shell is comprised of three broken segments in the highest frequency band (170-231\,MHz). The largest segment spans a circular angle of approximately 90$^{\circ}$ in the south-east, and smaller arcs ($\sim$45$^{\circ}$) of emission are present in the north-east and north-west (see Figure\,\ref{fig:MWAindividual}).



{We noted the presence of a point source within the shell of \gtt, and searched for information on this source in other surveys. The 1st Alternative Data Release of the Tata Institute for Fundamental Research Giant Metrewave Radio Telescope Sky Survey \citep[TGSS-ADR1; ][]{Intema:2017} shows an unresolved source with total flux density $S=805\pm80$\,mJy at 150\,MHz, while the National Radio Astronomy Observatory Very Large Array Sky Survey \citep[NVSS; ][]{1998AJ....115.1693C} measures the same source with $S=85.4\pm3.1$\,mJy at 1400\,MHz. NVSS has the best astrometry of these surveys and gives a position of $\mathrm{RA}=18^\mathrm{h}32^\mathrm{m}32.25^\mathrm{s},~\mathrm{dec}=-08^\mathrm{d}38^\mathrm{m}54.5^\mathrm{s}$ for the source.
Combining the two flux density measurements yields a source with spectral index $\alpha=-1.00\pm0.05$. As the source is unresolved and has a steep spectral index, we do not believe it to be part of the shell itself. It is $3'$ from the centroid of \gtt, and could potentially be a radio-bright pulsar, potentially mis-aligned and thus not appearing in existing pulsar surveys of the region.}

Alternatively, the source could be a radio galaxy with chance coincidence along the line-of-sight. \cite{2017MNRAS.464.1146H} surveyed 24,402~square degrees of the sky over 72--231\,MHz and created a catalogue of 307,456~radio sources at 200\,MHz, with an estimated completeness of 98\,\% at 600\,mJy, the flux density of the point source at this frequency. There are 27,628~sources with $S_\mathrm{200MHz}>0.6$\,Jy, reducing to 1,163 for sources where $\alpha$ is also less than $-1.1$. Therefore, the chance coincidence of any random radio galaxy lying anywhere within the shell is 13\,\%. This reduces to $\approx0.6$\,\% for any source lying within a radius of $3'$ from the shell centroid, or a steep-spectrum source lying anywhere within the shell. The chance of a steep-spectrum radio galaxy of the required flux density lying within $3'$ of the centroid is just 0.03\,\%.

\newtext{The \textit{Wide-field Infrared Survey Explorer} \citep[\textit{WISE};][]{2010AJ....140.1868W} surveyed the sky at 3.4, 4.6, 12, and 22\,$\mu$m, producing relatively high-resolution and high-sensitivity IR images of the entire sky. Radio galaxies selected at low frequencies and high flux densities ($S_\mathrm{150 MHz}\gtrsim10$\,mJy) are typically active galactic nuclei \citep{2008MNRAS.388.1335W} and are usually visible at 3.4 and 4.6$\mu$m unless they lie at high redshift ($z>1$). There is no IR counterpart to the radio source in any of the \textit{WISE} bands, further lending weight to the idea that the source is an associated compact remnant.}

\begin{figure*}
\centering
\includegraphics[width=0.85\textwidth]{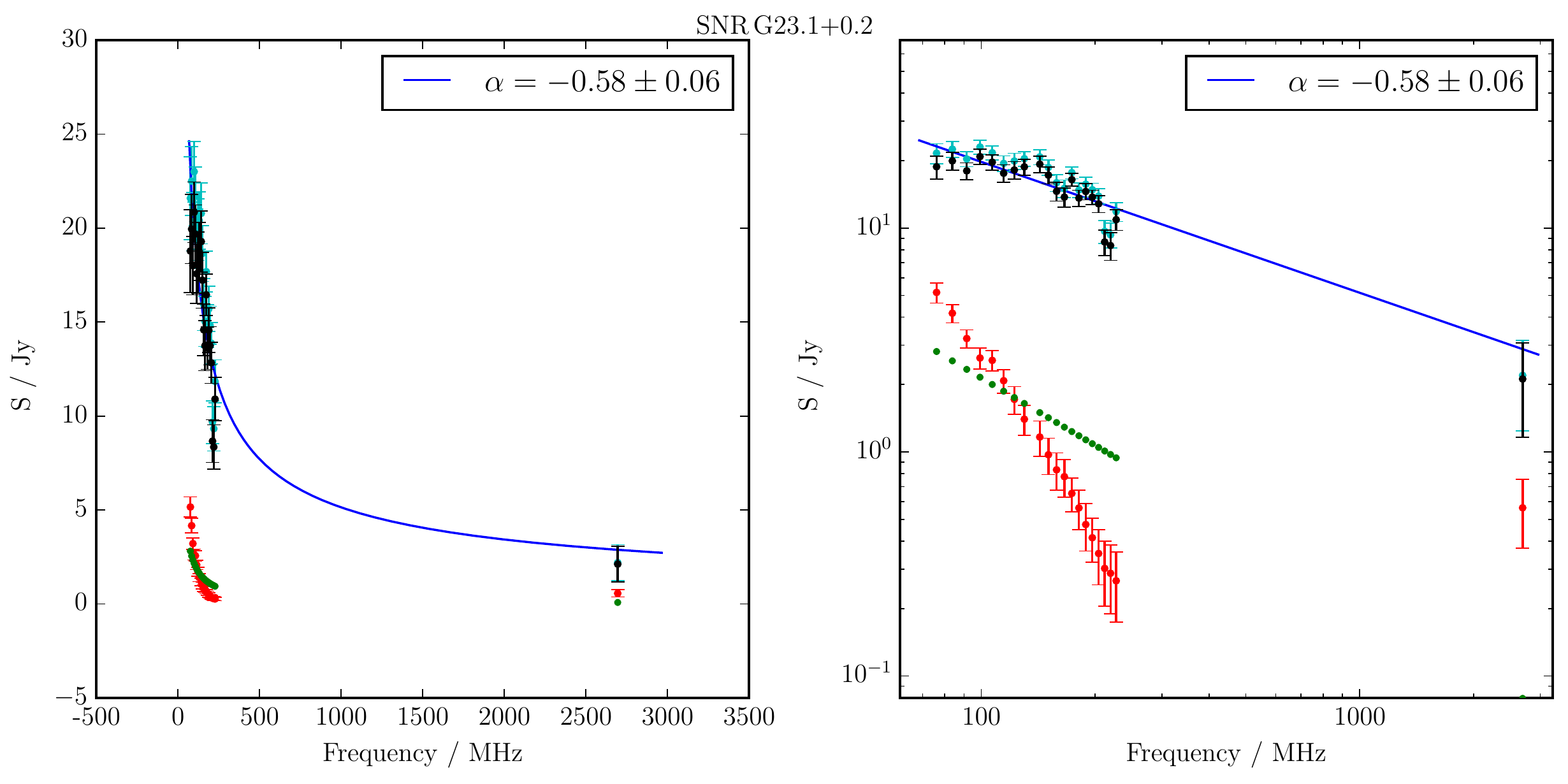}
\caption{Radio continuum spectra of \gtt. Linear and log scales are shown on the left and right, respectively. \gtt\ GLEAM and Effelsberg 2.695\,GHz flux density measurements are in cyan before the contaminating radio source (green) is subtracted, and black after; the background data points are in red, and the power law trend line fit is in blue.}
 \label{fig:MWAspec}
\end{figure*}

\begin{figure*}
\centering
\includegraphics[trim={0 0 0 0}, clip,width=0.85\textwidth]{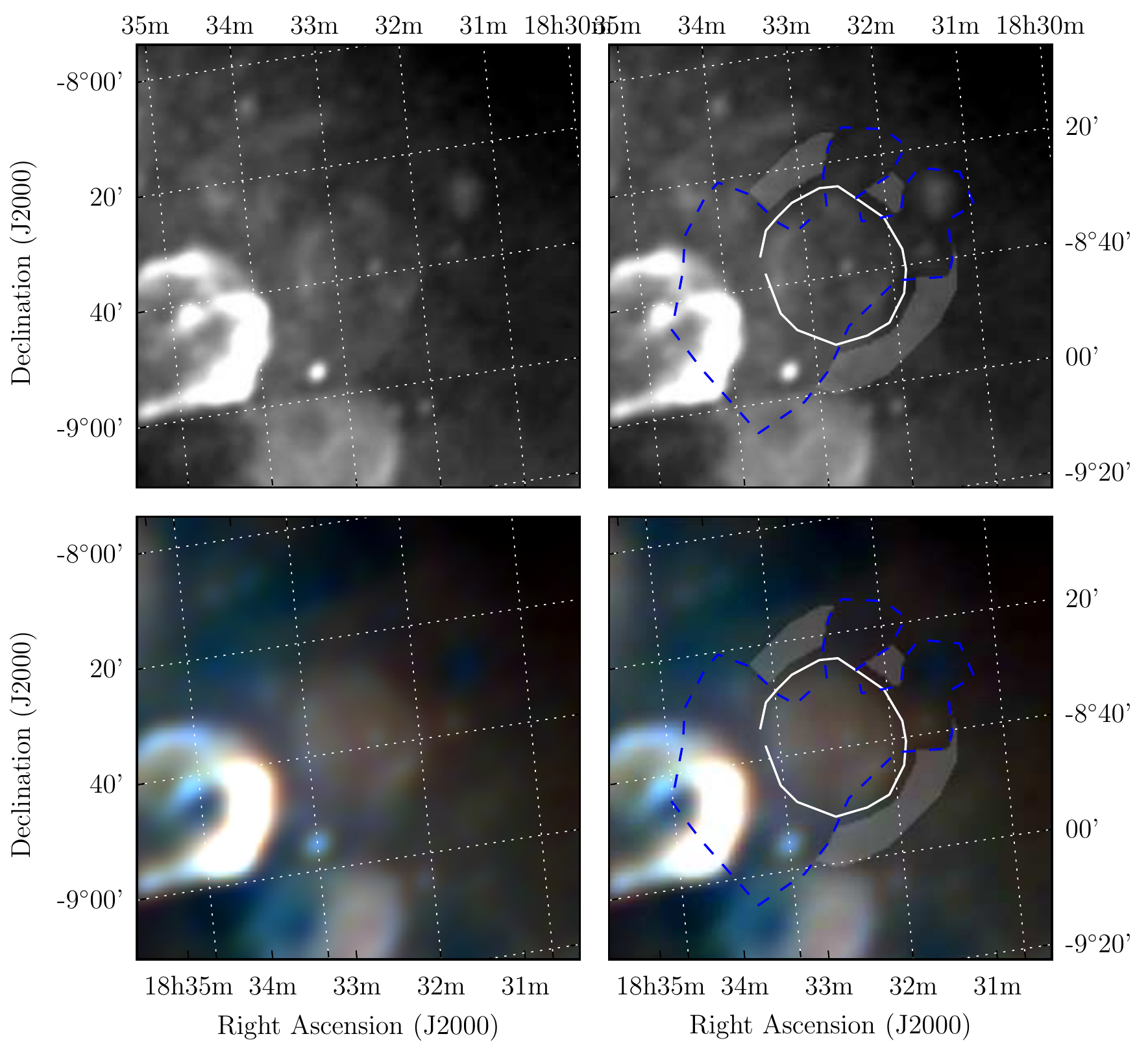}\\
\caption{The MWA radio continuum images used for source extraction. The top two images show the 170--231\,MHz data while the lower two images show RGB cubes of 72--103\,MHz (R), 103--134\,MHz (G), and 139--170\,MHz (B). The colour scales for the GLEAM RGB cube are 3.5--14.6, 1.5--7.7, and 0.6--4.2\,Jy\perbeam for R, G, and B, respectively.  In the {rightmost} images, the white polygon indicates the region used for the \gtt\ flux density estimation, the blue dashed polygon indicates a region excluded from the flux density estimation, and the shaded region indicates the region used to examine the background flux density. 
 \label{fig:MWAspecExtraction}}
\end{figure*}

The data points were estimated by extracting the total integrated flux density within a polygon that tightly enclosed the \gtt\ emission (see Figure\,\ref{fig:MWAspec}). A surrounding region, indicated {in Figure\,\ref{fig:MWAspecExtraction},} was used to subtract the flux of the background flux density for each band, while areas containing probable contamination nearby W41 and \textsc{Hii} regions were excluded from the calculation. The uncertainties on the flux density measurements are largely driven by the flux calibration accuracy of the data, about 8\,\% in this region. For more details about this process applied to a large sample of new and known SNRs, see \citet{Hurley-Walker_cand:2019} and \cite{Hurley-Walker_new:2019}.

{We obtained a Bonn 11-cm Survey single dish 2.695\,GHz} image of the region and measured the total flux density of the SNR, integrating over the same area and using the same region as used for background subtraction of the GLEAM data. The signal-to-noise of the SNR is much lower, due to the higher background level from the single dish measurement, and the lower brightness of the SNR at higher frequencies due to its non-thermal spectrum.

As shown in Figure\,\ref{fig:MWAspec}, we subtracted the point source from the GLEAM and Effelsberg data, and calculated the 72--2695\,MHz radio spectral index\footnote{Defined as $S_{\nu}\propto\nu^{\alpha}$} as $\alpha=-0.58\pm0.06$, which is within the typical range of SNRs \citep[$-$0.3 to $-$0.8][]{Bozzetto:2017} for a wide range of diameters and ages (5-100\,pc and 200-10$^6$\,yr, respectively)). Table\,\ref{tab:MWAflux} lists the radio continuum integrated flux densities of \gtt and the contaminating point source at each frequency.

\begin{table}
\centering
\begin{tabular}{|c|c|c|c|}
\hline
$\nu$ Frequency & Total flux density & Predicted point source flux density & \gtt~ flux density after source-subtraction\\ 
(MHz)  & (Jy) & (Jy) & (Jy) \\
\hline
$  76$ & $21.6\pm2.2$ & $2.81\pm0.14$ & $18.8\pm2.2$ \\
$  84$ & $22.5\pm1.8$ & $2.55\pm0.13$ & $19.9\pm1.8$ \\
$  92$ & $20.3\pm1.6$ & $2.34\pm0.12$ & $18.0\pm1.6$ \\
$  99$ & $23.0\pm1.6$ & $2.16\pm0.11$ & $20.8\pm1.6$ \\ 
$ 107$ & $21.7\pm1.6$ & $2.00\pm0.10$ & $19.7\pm1.6$ \\
$ 115$ & $19.4\pm1.6$ & $1.87\pm0.09$ & $17.6\pm1.6$ \\
$ 122$ & $19.9\pm1.6$ & $1.75\pm0.09$ & $18.2\pm1.6$ \\
$ 130$ & $20.4\pm1.5$ & $1.65\pm0.08$ & $18.8\pm1.5$ \\
$ 143$ & $20.8\pm1.6$ & $1.50\pm0.07$ & $19.3\pm1.6$ \\
$ 150$ & $18.7\pm1.5$ & $1.42\pm0.07$ & $17.2\pm1.5$ \\ 
$ 158$ & $15.9\pm1.4$ & $1.35\pm0.07$ & $14.6\pm1.4$ \\
$ 166$ & $15.0\pm1.3$ & $1.29\pm0.06$ & $13.7\pm1.3$ \\
$ 173$ & $17.7\pm1.1$ & $1.23\pm0.06$ & $16.4\pm1.1$ \\
$ 181$ & $14.8\pm1.1$ & $1.18\pm0.06$ & $13.6\pm1.1$ \\
$ 189$ & $15.7\pm1.2$ & $1.13\pm0.06$ & $14.6\pm1.2$ \\
$ 196$ & $14.8\pm1.0$ & $1.09\pm0.05$ & $13.7\pm1.0$ \\
$ 204$ & $13.9\pm1.1$ & $1.05\pm0.05$ & $12.8\pm1.1$ \\
$ 212$ & $9.7\pm1.1$ & $1.01\pm0.05$ & $8.7\pm1.1$ \\ 
$ 220$ & $9.3\pm1.2$ & $0.97\pm0.05$ & $8.3\pm1.2$ \\
$ 227$ & $11.8\pm1.1$ & $0.94\pm0.05$ & $10.9\pm1.2$ \\
$2695$ & $2.2\pm1.0$ & $0.08\pm0.01$ & $2.1\pm1.0$ \\
\hline
\end{tabular}
\caption{Radio flux densities derived from MWA data for \gtt. 
 \label{tab:MWAflux}}
\end{table}

\subsection{\hje\ and a Search for a Gas Association}\label{sec:Gamma}\label{sec:Gas}
\begin{figure}
\centering
\includegraphics[width=0.49\textwidth]{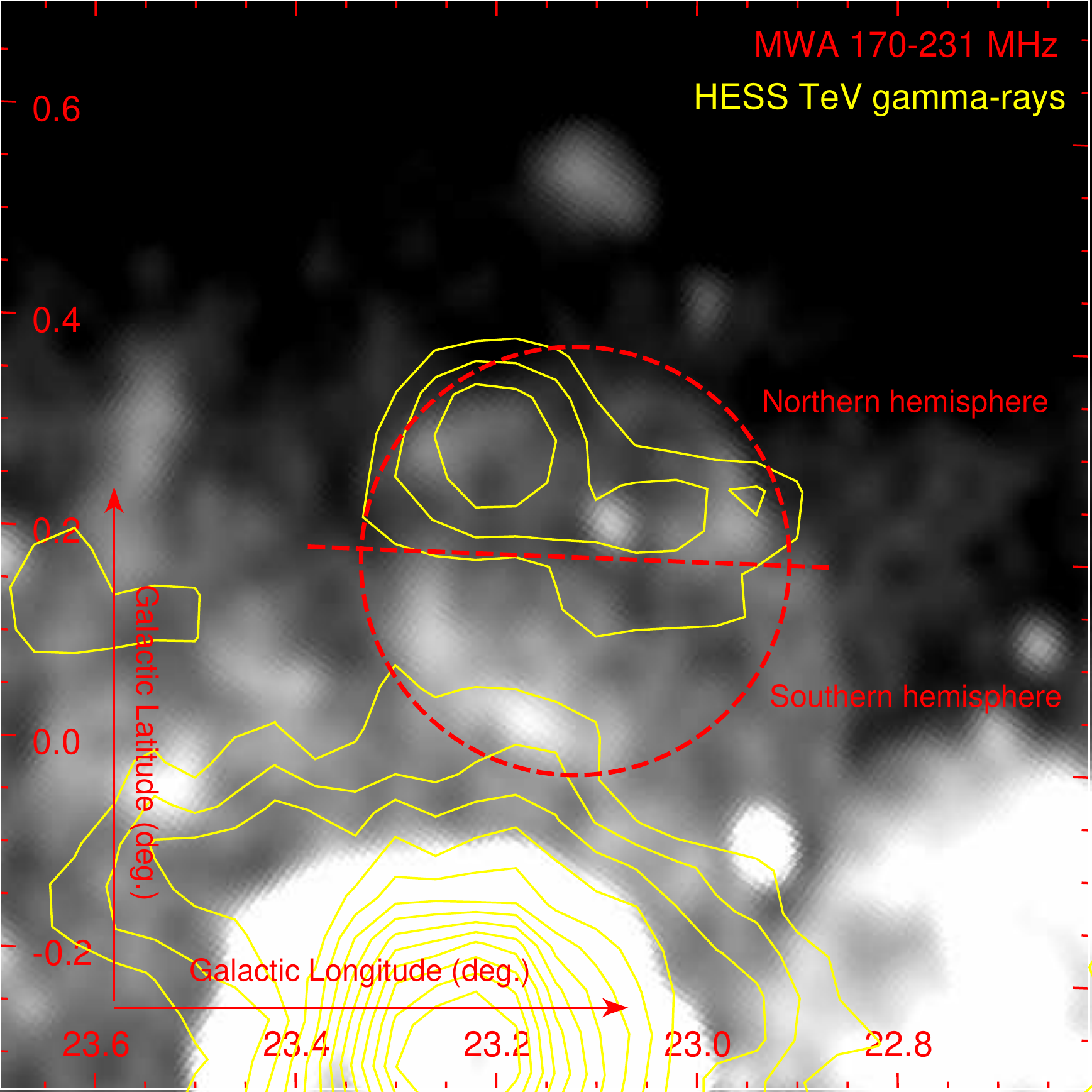}
\caption{Murchison Widefield Array 130-231\,MHz radio continuum image of SNR candidate \gtt\ with HESS TeV Gamma-ray significance contours overlaid (contour increment is 1$\sigma$, beginning at 3$\sigma$). A red circle indicates the proposed location of \gtt . 
 \label{fig:MWA_wCircAndHESS}}
\end{figure}

Figure\,\ref{fig:MWA_wCircAndHESS} shows the HESS Galactic Plane Survey TeV gamma-ray map overlaid on MWA radio continuum data. The unidentified gamma ray source, \hje , is coincident with the northern region of the new SNR candidate. As gamma-ray emission associated with SNR shells is not uncommon \citep[e.g. see][]{Abdalla:2018_SNRpop}, we propose that \hje\ {is likely} associated with \gtt . We further argue that such a coincidence reinforces the candidacy status of \gtt\ as a new Galactic SNR.\footnote{Indeed the independent identification of \gtt\ in two radio continuum data-sets \citep[in this MWA study, and in THOR data,][]{Anderson:2017_THORsnrs} is already strong evidence in favour of the shell-like feature.}

A search for molecular gas towards \gtt\ can shed light on SNR nature in three ways.
{Firstly, among young ($\sim$10$^{3}$\,yr) SNRs in the Sedov-Taylor phase, evolution is often seen to be occurring in an inhomogeneous, clumpy medium \citep[e.g.][]{Moriguchi:2005,Zirakashvili:2010,Fukui:2012,Inoue:2012,2013MNRAS.428.1980D,Maxted:2012,Maxted:2013rxj,Maxted:2013ctb,Fukuda:2014,Fukui:2017,Sano:2017_rcw86,Maxted:2018_HESSJ1731,Maxted:2018_HESSJ1534,Maxted:2018_Vela,Sano:2018_RCW86}. It follows that a search for clumps towards the gamma-ray source \hje\ can provide an insight into its age, environment, and possible relation to \gtt.}

{Secondly, gamma-ray emission is observed towards SNRs that are middle-aged ($\sim$10$^{4}$\,yr) and in the radiative phase. The gamma-ray emission seems to result from a lingering population of CR protons that diffuse into nearby molecular gas \citep[e.g. SNR W28;][]{Arikawa:1999,Aharonian:w28,Gabici:2007,Gabici:2008,Gabici:2009,Gabici:2010,Nicholas:2011,Nicholas:2012,Maxted:2016a,Maxted:2017w28}. A search for molecular clouds corresponding to the \hje\ gamma-ray emission allows us to test if a similar scenario is occurring for \gtt. In our investigation, this `nearby cloud' scenario might be difficult to distinguish from an `embedded clump' scenario in the case that {a nearby inhomogeneous cloud} lies adjacent to the SNR in the line of sight.}

Finally, the identification of a void or cavity in molecular gas corresponding to similar dimensions to \gtt\ provides a plausible gas association in a progenitor wind-blown bubble scenario. 

{In the following sections, we examine} FUGIN CO, $^{13}$CO and C$^{18}$O(1-0) data in a search for both dips in molecular gas corresponding to the new SNR candidate, and gas coincident to the unidentified gamma-ray source \hje, which might be candidate embedded clumps or nearby clouds illuminated by CRs. {We also perform a search for the SNR candidate in other frequencies.}

\subsubsection{Galactic Structure towards \gtt}\label{ssec:GalStruc}
\begin{figure*}
\centering
\includegraphics[width=1.05\textwidth]{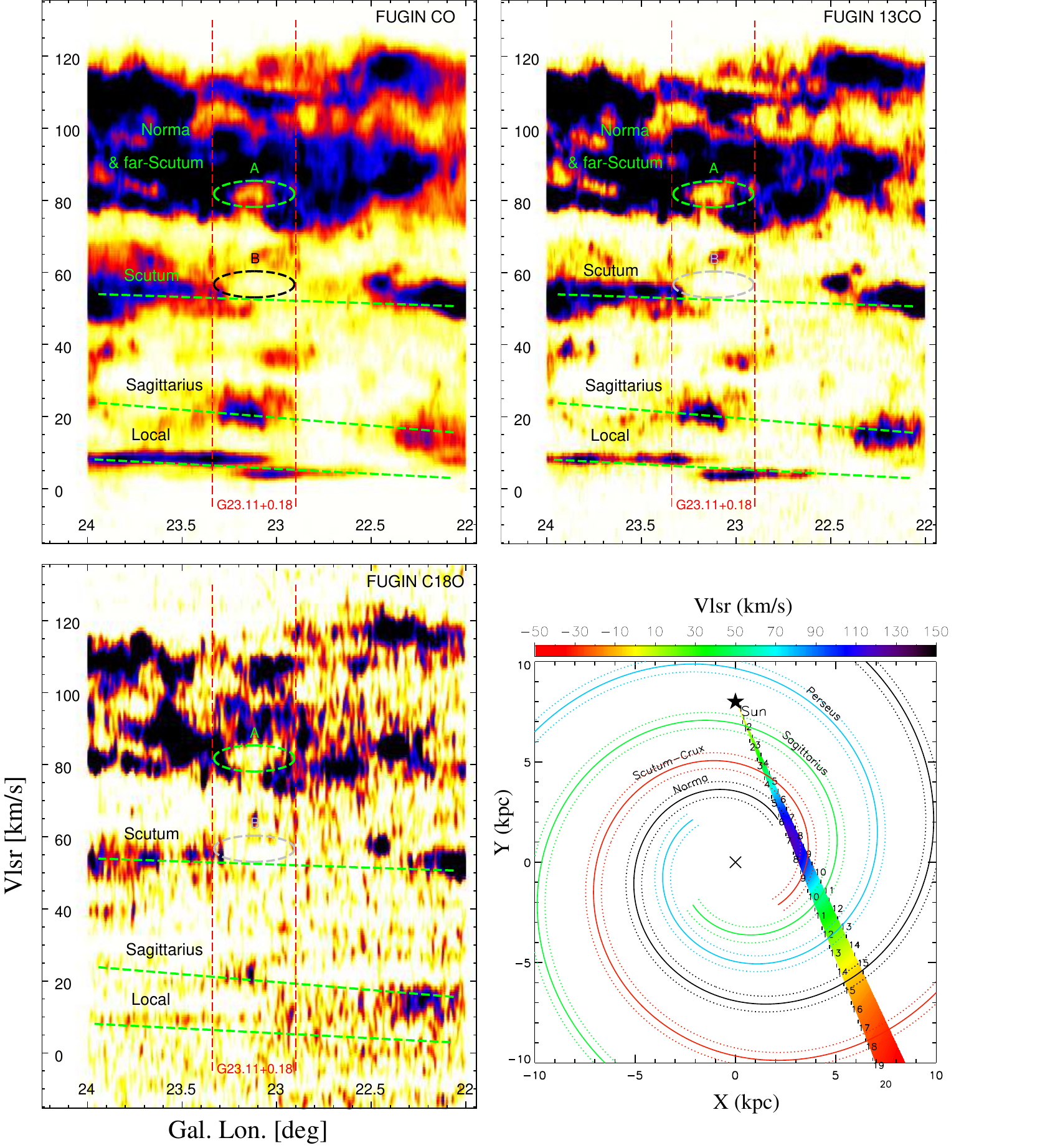}
\caption{$^{12}$CO(1-0), $^{13}$CO(1-0) and C$^{18}$O(1-0) emission as a function of Galactic longitude and line of sight velocity (top-left, top-right and bottom-left, respectively). Data have been integrated in the Galactic latitude dimension between 0 and 0.4 degrees, corresponding to the candidate SNR \gtt . Darkness in colour indicates increasing intensity. Red dashed lines indicate the Longitudinal extent of \gtt , green dashed lines indicate the approximate locations of Galactic arms, as extrapolated from Figure\,8 of \citet{Umemoto:2017_FUGIN1}. Circles indicate two void-candidates, Void\,A and Void\,B. On the bottom-right is a schematic of the Galactic arm model in \citet{Vallee:2016}, where colour is used to indicate kinematic velocity {in the direction of this field.}. 
\label{fig:PVplots_lon}}
\end{figure*}

In Figure\,\ref{fig:PVplots_lon} we display longitude-velocity plots of the \gtt\ region. The most intense structures in the three CO isotopologue tracers are between 75 and 120\,km~s$^{-1}$. These high-velocity components likely correspond to molecular gas near the Norma-arm tangent and the far-side of the Scutum-Crux arm. Three additional line of sight components are indicated: Scutum-Crux-arm (near), Sagittarius arm and local molecular gas. The attribution of names for these structures are extrapolated from \citet{Umemoto:2017_FUGIN1} and the Milky Way schematic in Figure\,\ref{fig:PVplots_lon}. Although it is possible that components of the far side of the Sagittarius and Perseus arm are present in position-velocity plots, we do not attempt make further inferences about Galactic structure from this data-set, and instead simply focus on associations for \gtt .

\subsubsection{Possible gas signatures of \gtt}\label{ssec:gas_sig}
Two void-like structures were identified by eye in velocity-space, and are indicated in Figure\,\ref{fig:PVplots_lon}. Void\,A is between 80 and 90\,km~s$^{-1}$ within molecular gas most-likely attributable to the Norma arm. Void\,B is between 50 and 60\,km~s$^{-1}$ within molecular gas most-likely attributable to the Scutum arm.

Void A is completely encircled in $^{12}$CO(1-0) and $^{13}$CO(1-0) emission in velocity-space, while a $\sim$0.1\,degree gap can be observed in the low-velocity side of C$^{18}$O emission. Void\,B is more sparse with $^{12}$CO, $^{13}$CO and C$^{18}$O(1-0) emission on the higher-longitude side and a central clump on the high velocity side. $^{12}$CO(1-0) and $^{13}$CO(1-0) emission partially envelops Void\,B on both sides in velocity-space. We consider these two void-like structures as possible associations for \gtt .

\begin{figure*}
\centering
\includegraphics[width=0.33\textheight]{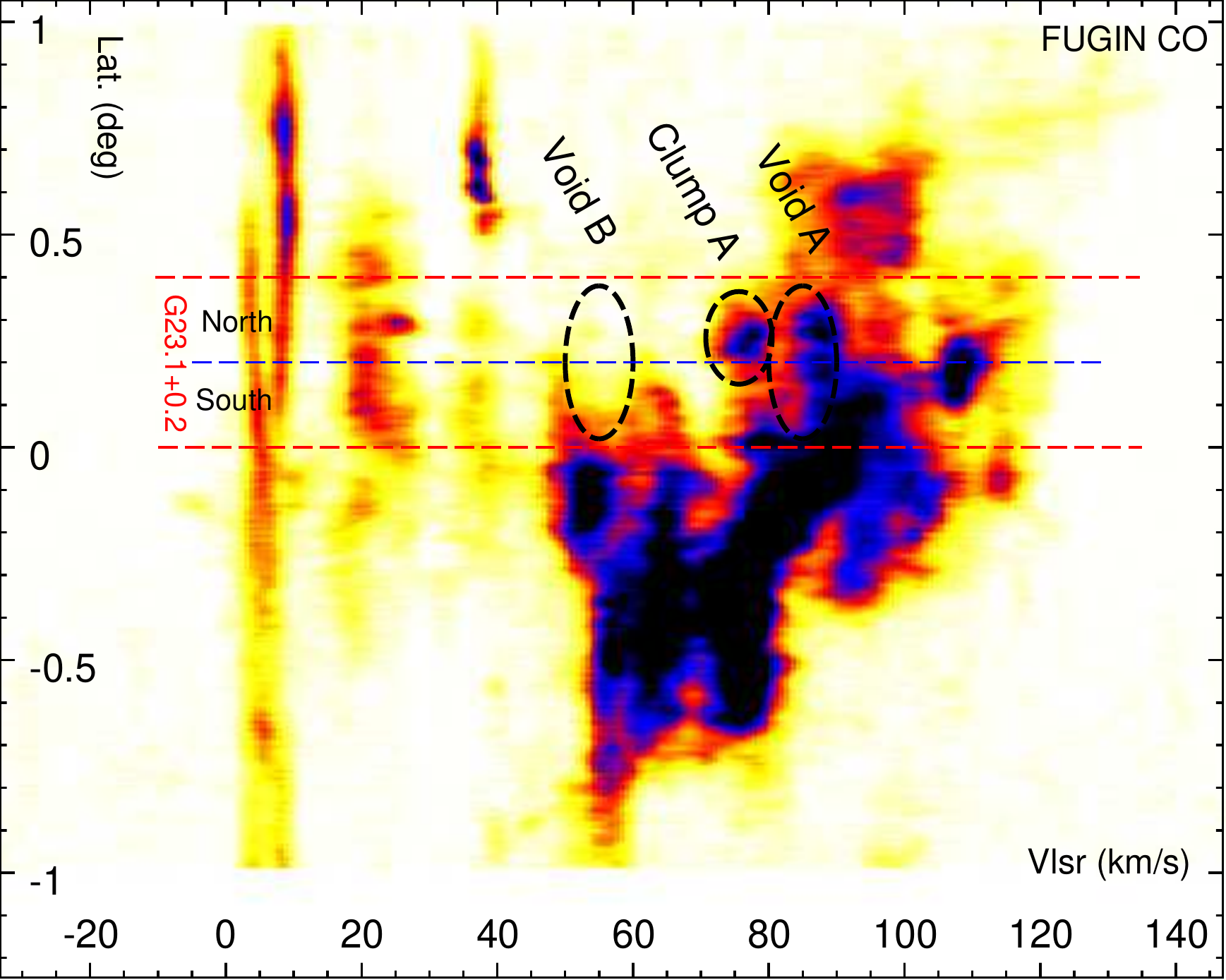}\\
\includegraphics[width=0.33\textheight]{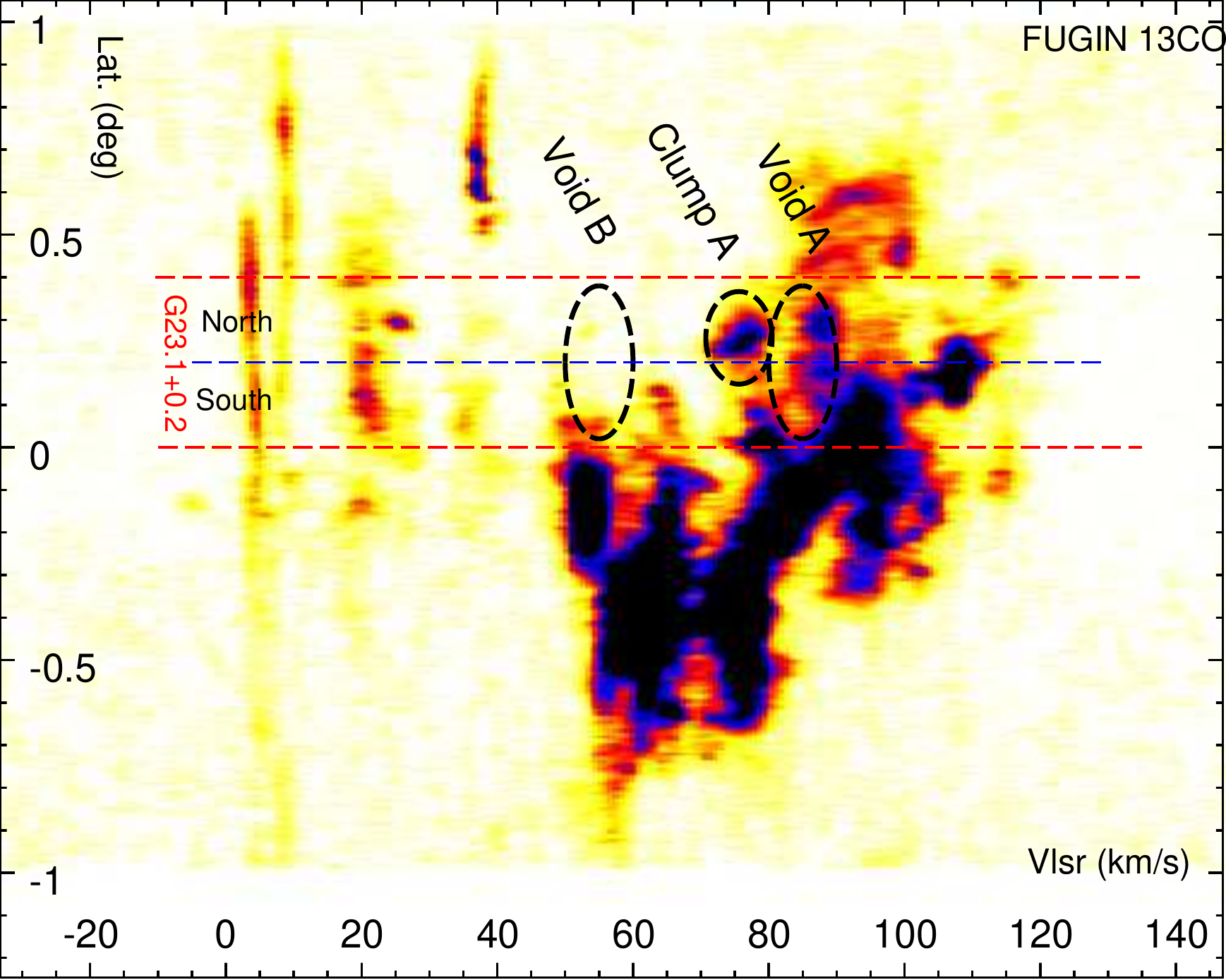}\\
\includegraphics[width=0.33\textheight]{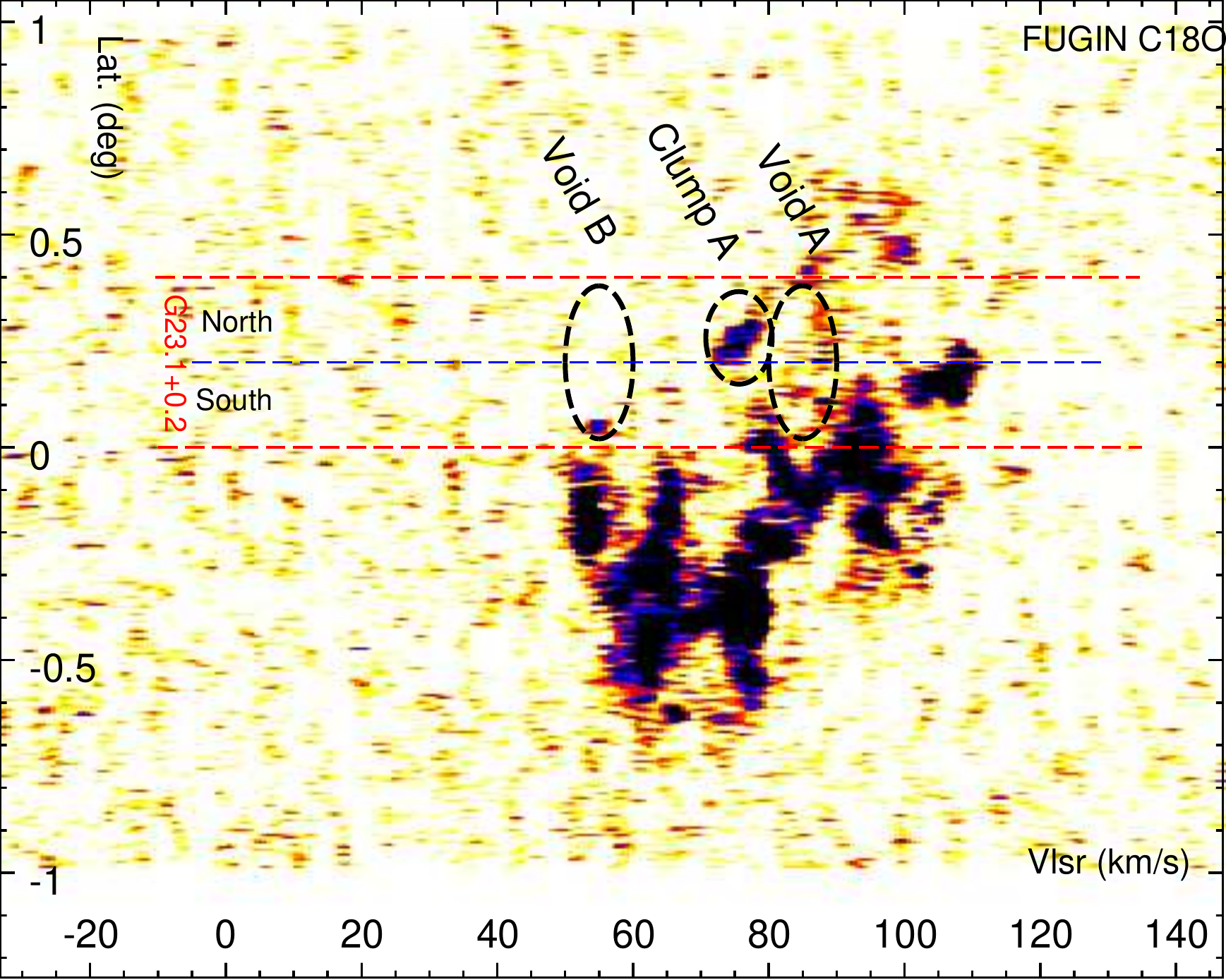}
\caption{$^{12}$CO(1-0), $^{13}$CO(1-0) and C$^{18}$O(1-0) emission as a function of Galactic latitude and line of sight velocity. Data have been integrated in the Galactic longitude dimension between 22.9 and 23.34 degrees, corresponding to the candidate SNR \gtt. Darkness in colour indicates increasing intensity. Red dashed lines indicate the Longitudinal extent of \gtt, while a blue line indicates the mid-latitude point of the SNR. Two \gtt\ void candidates, Void\,A and B, and Clump\,A are indicated with ellipses. 
 \label{fig:PVplots_lat}}
\end{figure*}

In Figure\,\ref{fig:PVplots_lat} we display latitude-velocity plots of the \gtt\ region. 
Figure\,\ref{fig:PVplots_lat} shares some velocity characteristics in common with Figure\,\ref{fig:PVplots_lon}, i.e. structures of Norma, Scutum, Sagittarius and local gas components are likely present.

Imaging the CO data in the latitudinal dimension allows us to investigate correspondence with the unidentified gamma-ray source \hje . Since \hje\ appears in the northern half of \gtt, we search for velocity ranges that contain significantly more gas in the northern hemisphere of \gtt\ when compared to the southern hemisphere. This analysis carries an assumption that the gamma-ray emission should correlate with gas, as motivated by aforementioned known cases of gas/gamma-ray correlation caused by CRs interacting with molecular clouds (see Section\,\ref{sec:Gamma}).

From Figure\,\ref{fig:PVplots_lat}, one velocity range stands out as having significantly more emission in the north relative to the south: 70-80\,km~s$^{-1}$. This feature is most prominent in C$^{18}$O(1-0) emission, which highlights the most \newtext{CO(1-0) optically-thick components}. We indicate the C$^{18}$O(1-0) component as `Clump\,A' in Figure\,\ref{fig:PVplots_lat}.

From Figure\,\ref{fig:PVplots_lon} and \ref{fig:PVplots_lat}, we have flagged three velocity ranges for us to examine for potential association: two wind-blown void candidates for \gtt , Void\,A and B (80-90\,km~s$^{-1}$ and 50-60\,km~s$^{-1}$, respectively), and one clump association for \hje\ (Clump\,A, 70-80\,km~s$^{-1}$).

\begin{figure*}
\centering
\includegraphics[width=0.9\textwidth]{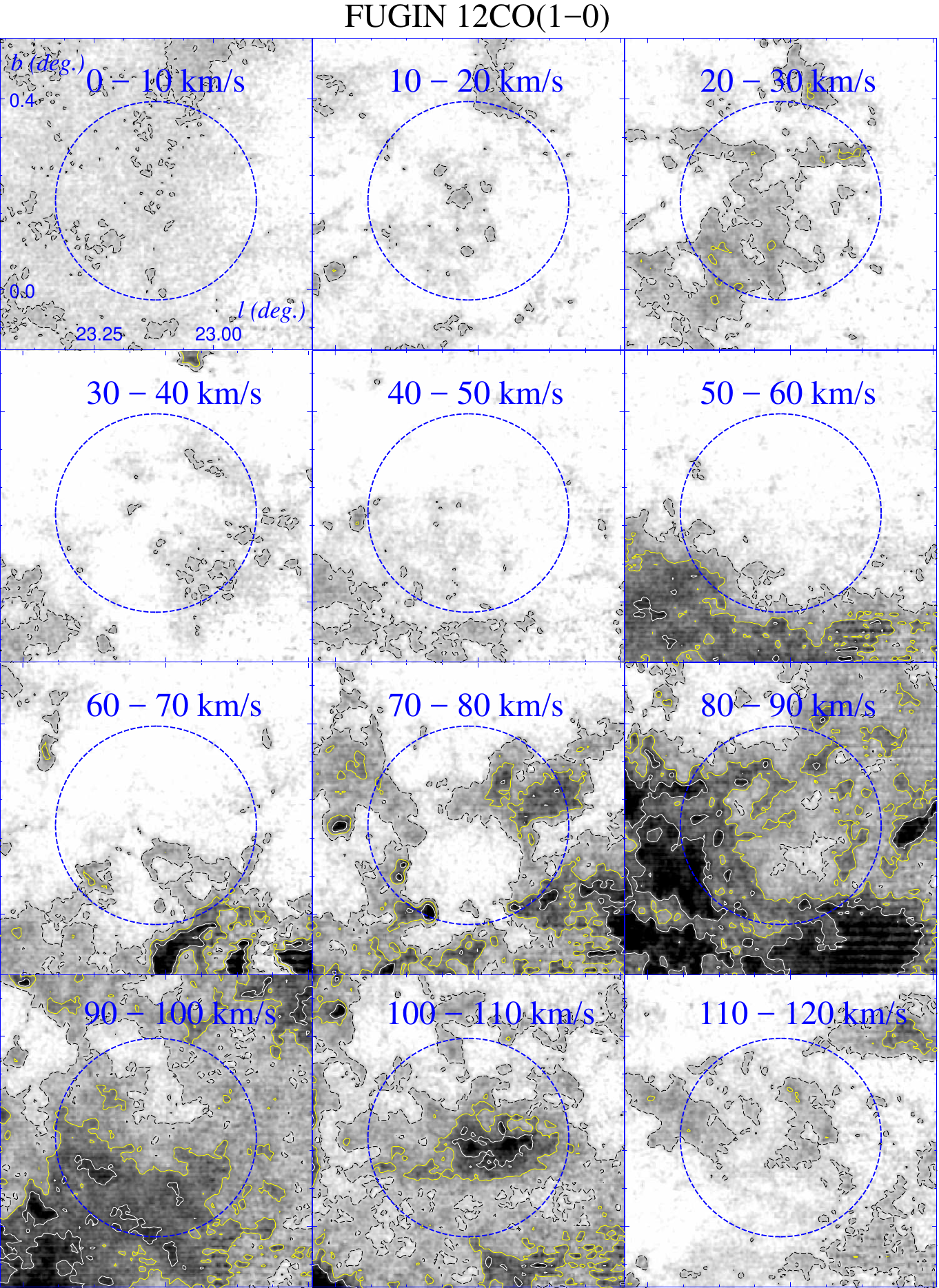}
\caption{FUGIN $^{12}$CO(1-0) integrated intensity images \citep{Umemoto:2017_FUGIN1}. {Darkness in colour indicates increasing intensity. Contour levels showing 20, 40 and 60\,K\,km\,s$^{-1}$ are overlaid (black-dashed, yellow and white, respectively).} Velocity integration ranges are indicated on each image. A circle indicates the approximate location of candidate SNR \gtt . 
\label{fig:12COslices}}
\end{figure*}

\begin{figure*}
\centering
\includegraphics[width=0.9\textwidth]{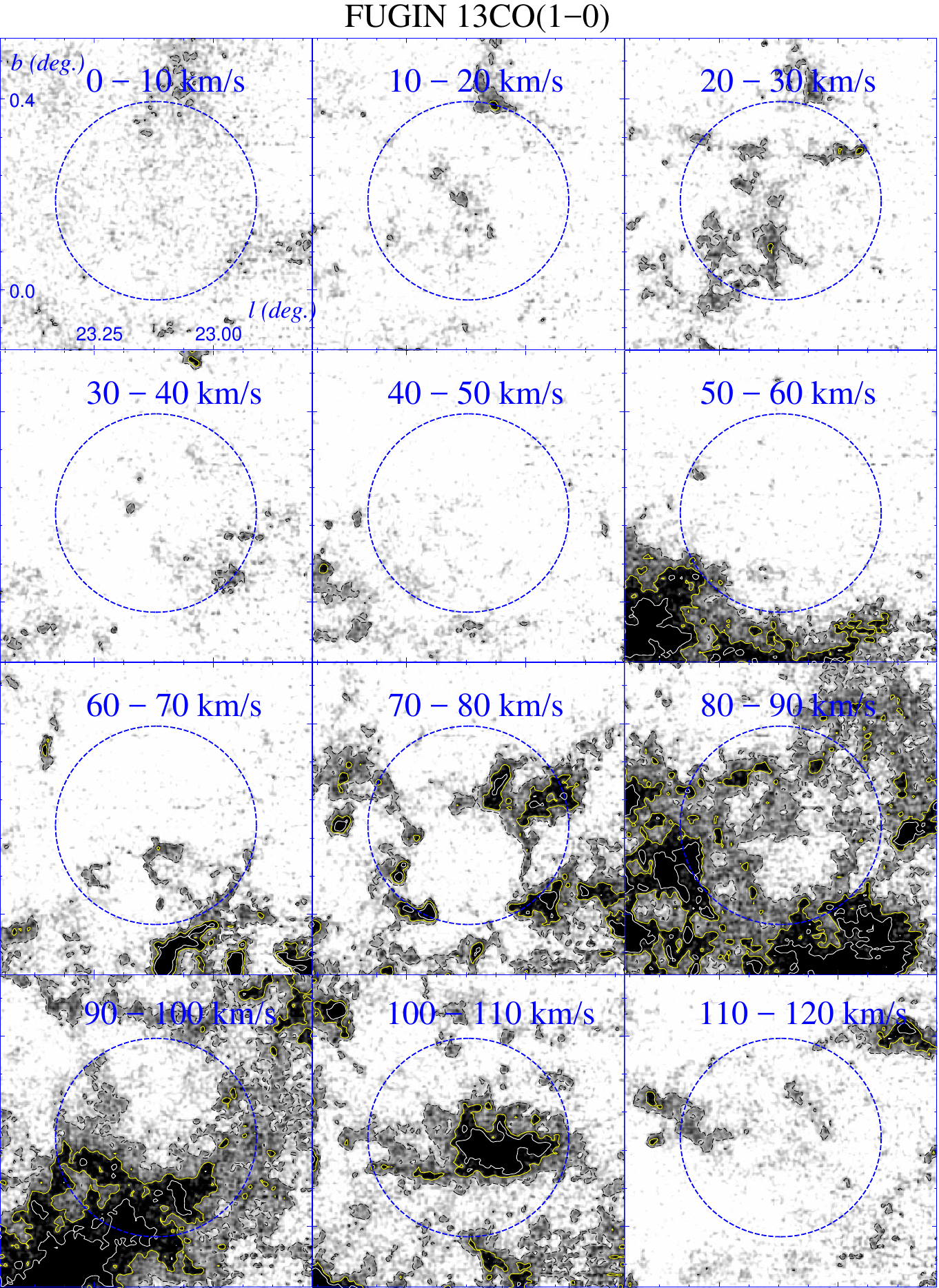}
\caption{FUGIN $^{13}$CO(1-0) integrated intensity images \citep{Umemoto:2017_FUGIN1}. {Darkness in colour indicates increasing intensity. Contour levels showing 6, 12 and 18\,K\,km~s$^{-1}$ are overlaid (black-dashed, yellow and white, respectively).} Velocity integration ranges are indicated on each image. A circle indicates the approximate location of candidate SNR \gtt . 
 \label{fig:13COslices}}
\end{figure*}

\begin{figure*}
\centering
\includegraphics[width=0.9\textwidth]{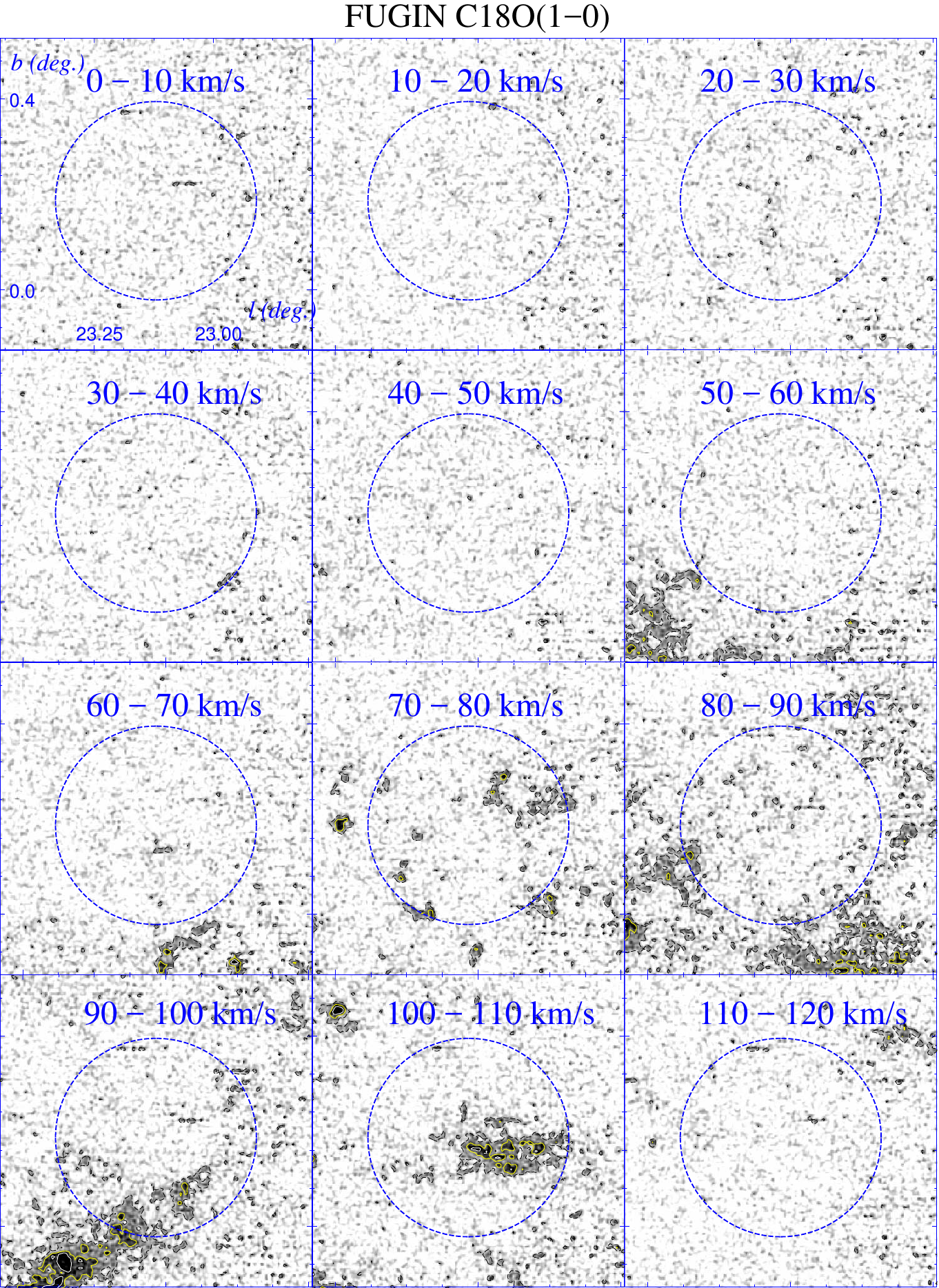}
\caption{FUGIN C$^{18}$O(1-0) integrated intensity images \citep{Umemoto:2017_FUGIN1}. {Darkness in colour indicates increasing intensity. Contour levels showing 3, 6 and 9\,K\,km~s$^{-1}$ are overlaid (black-dashed, yellow and white, respectively).} Velocity integration ranges are indicated on each image. A circle indicates the approximate location of candidate SNR \gtt.
 \label{fig:C18Oslices}
 }
\end{figure*}

Figure\,\ref{fig:12COslices}, \ref{fig:13COslices} and \ref{fig:C18Oslices} are 10\,km~s$^{-1}$-slices of CO(1-0), $^{13}$CO(1-0) and C$^{18}$O(1-0) emission, respectively. From these figures, the two candidate Void velocity ranges, are morphologically consistent with wind-blown cavity scenarios. The Void\,A velocity (80-90\,km~s$^{-1}$) shows a wall of molecular gas in CO and $^{13}$CO stretching from the north-east, down to the south-east and around to the south-west to encircle the SNR candidate over $\sim$180$^{\circ}$. Within the \gtt\ perimeter, and to the north-west, less intense, and clumpier emission exists. As expected from its relative rarity, C$^{18}$O(1-0) is generally less prominent than CO and $^{13}$CO(1-0) around Void\,A. There is, however, C$^{18}$O(1-0) emission towards the eastern, western and south-western edges of \gtt, indicating the presence of high CO optical-depth molecular clumps.

The Void\,B velocity (50-60\,km~s$^{-1}$) also exhibits a wall of molecular gas in CO and $^{13}$CO(1-0). Emission from these two diffuse molecular gas tracers surround the SNR candidates eastern, southern and south-western boundary -- approximately 120$^{\circ}$ of angular correspondence. C$^{18}$O(1-0) indicates the presence of optically-thick $^{12}$CO clumps on the south-eastern and southern \gtt\ boundary.

Clump\,A (70-80\,km~s$^{-1}$) is near the Void\,A velocity (80-90\,km~s$^{-1}$) and may form part of the same arm. The 70-80\,km~s$^{-1}$ velocity window in Figure\,\ref{fig:12COslices} and \ref{fig:13COslices} exhibits CO(1-0) and $^{13}$CO(1-0) emission overlapping with the north-west and east of \gtt, while a dip in emission in the south is bordered by emission at the southern edge of \gtt. C$^{18}$O emission is also prominent in the north-west quadrant of \gtt\ and around the south-east and south-west boundaries. 

The correspondence of the 70-80\,km~s$^{-1}$ CO emission with the \hje\ gamma-ray emission is incomplete in the north-east of \gtt, but this significantly improves by including the neighboring Void\,A velocity of 80-90\,km~s$^{-1}$ (see Figure\,\ref{fig:COthreeCol}). Physically, this can be justified under the assumption that the emission originates from the same arm, with both Void\,A and Clump\,A having an association with \gtt, particularly if local kinematic cloud motions of 5-10\,km/s exist. {The association is consistent with both a clumpy medium within the \gtt\ shell or gas immediately adjacent to the SNR in the line of sight. } 
{In Figure\,\ref{fig:COthreeCol}, a CO/$^{13}$CO clump is coincident with the TeV gamma-ray peak of \hje . From this, we suggest that the initial `point-like' classification of the source is simply reflecting a gamma-ray flux that follows gas distribution inside or near \gtt . } 

\subsection{Optical and Infrared Searches for \gtt}\label{sec:Infrared}
We were unable to find morphological signatures of the SNR candidate \gtt\ in H$\alpha$ images taken in 1999 as part of the SuperCOSMOS Sky Survey with the UK-Schmidt telescope \citep{Hambly:2001}. Furthermore, we did not find \gtt\ in Midcourse Space Experiment or Spizter-GLIMPSE infrared data \citep[][respectively]{Price:2001,Churchwell:2009}. 

\begin{figure*}
\centering
\includegraphics[width=0.49\textwidth]{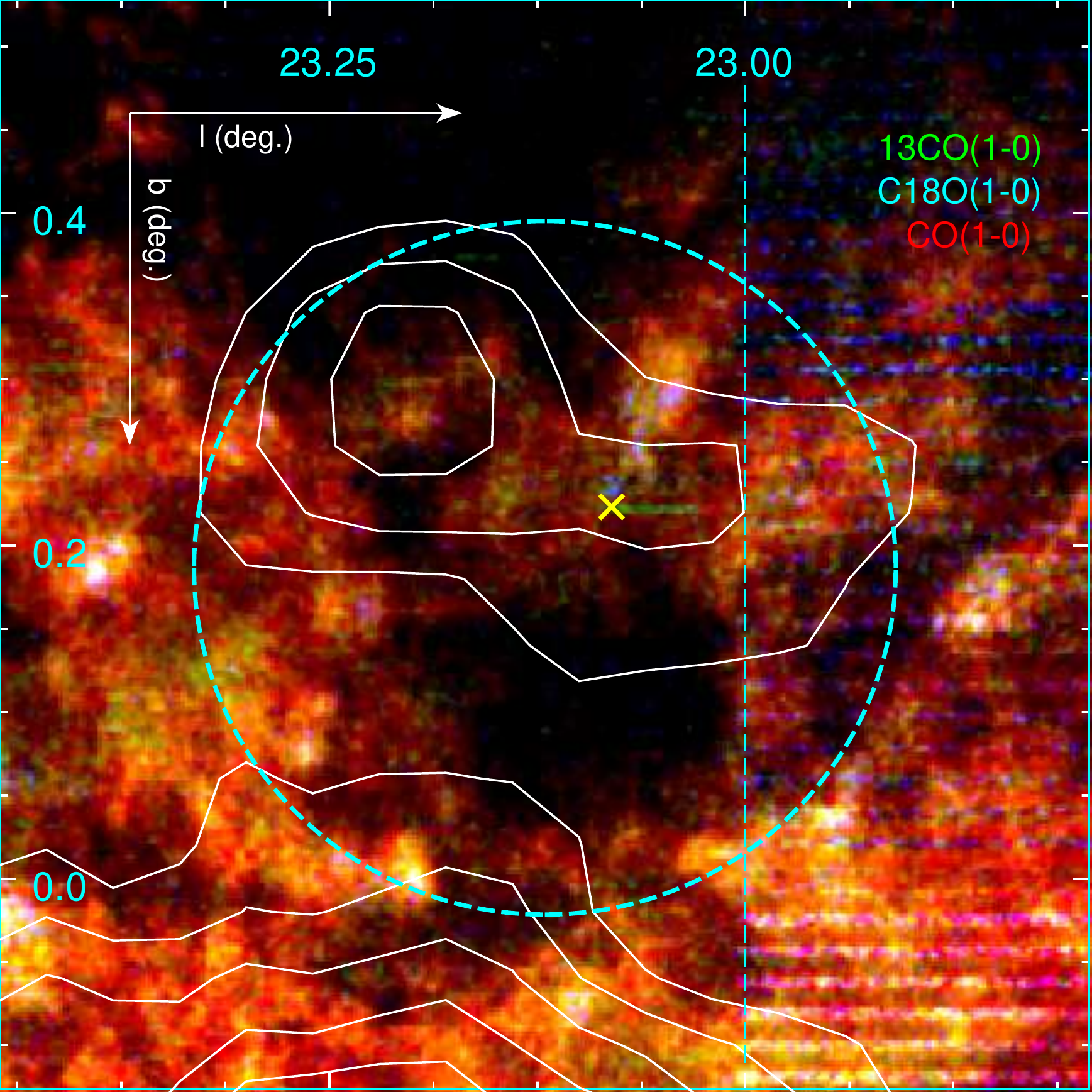}
\includegraphics[width=0.49\textwidth]{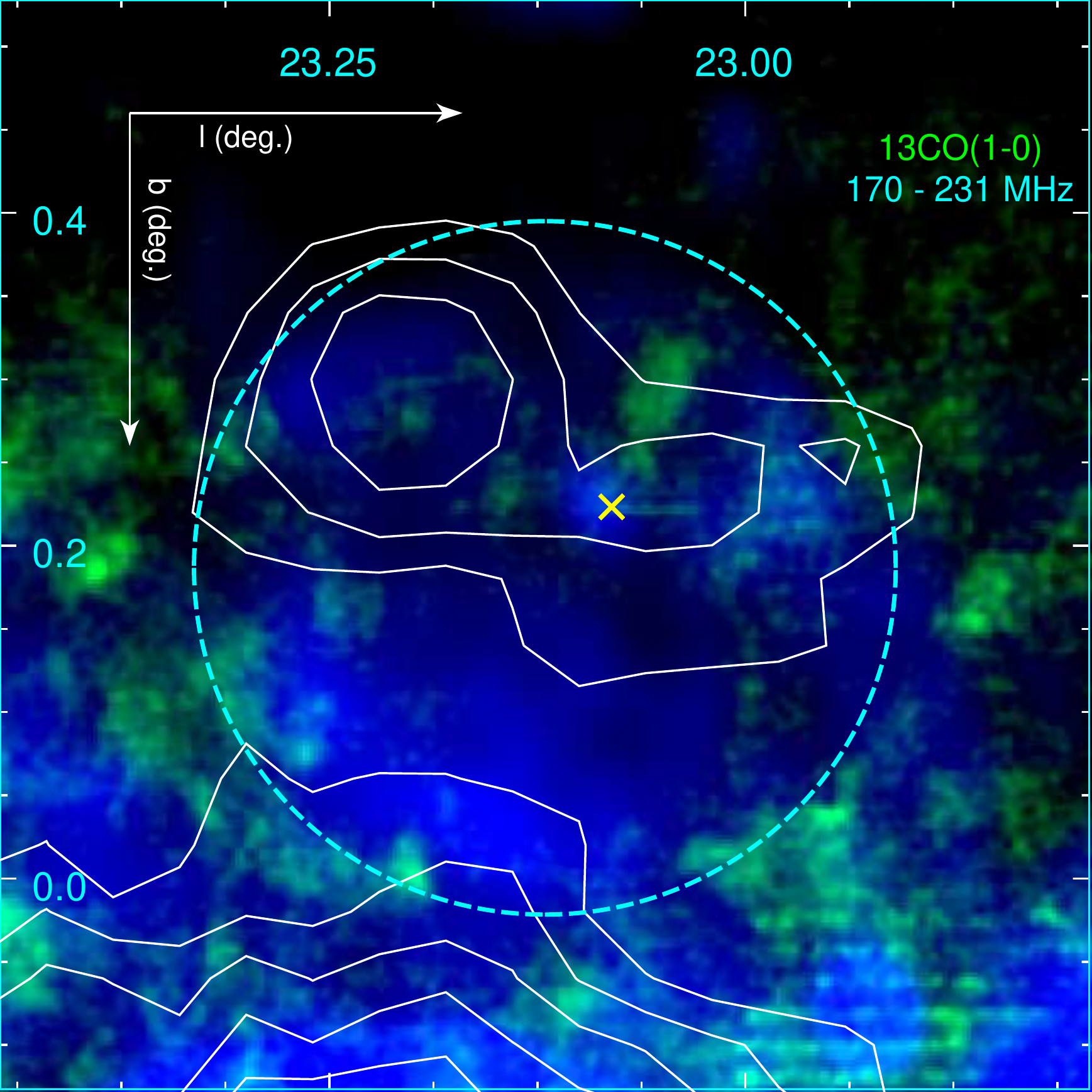}\\
\includegraphics[width=0.49\textwidth]{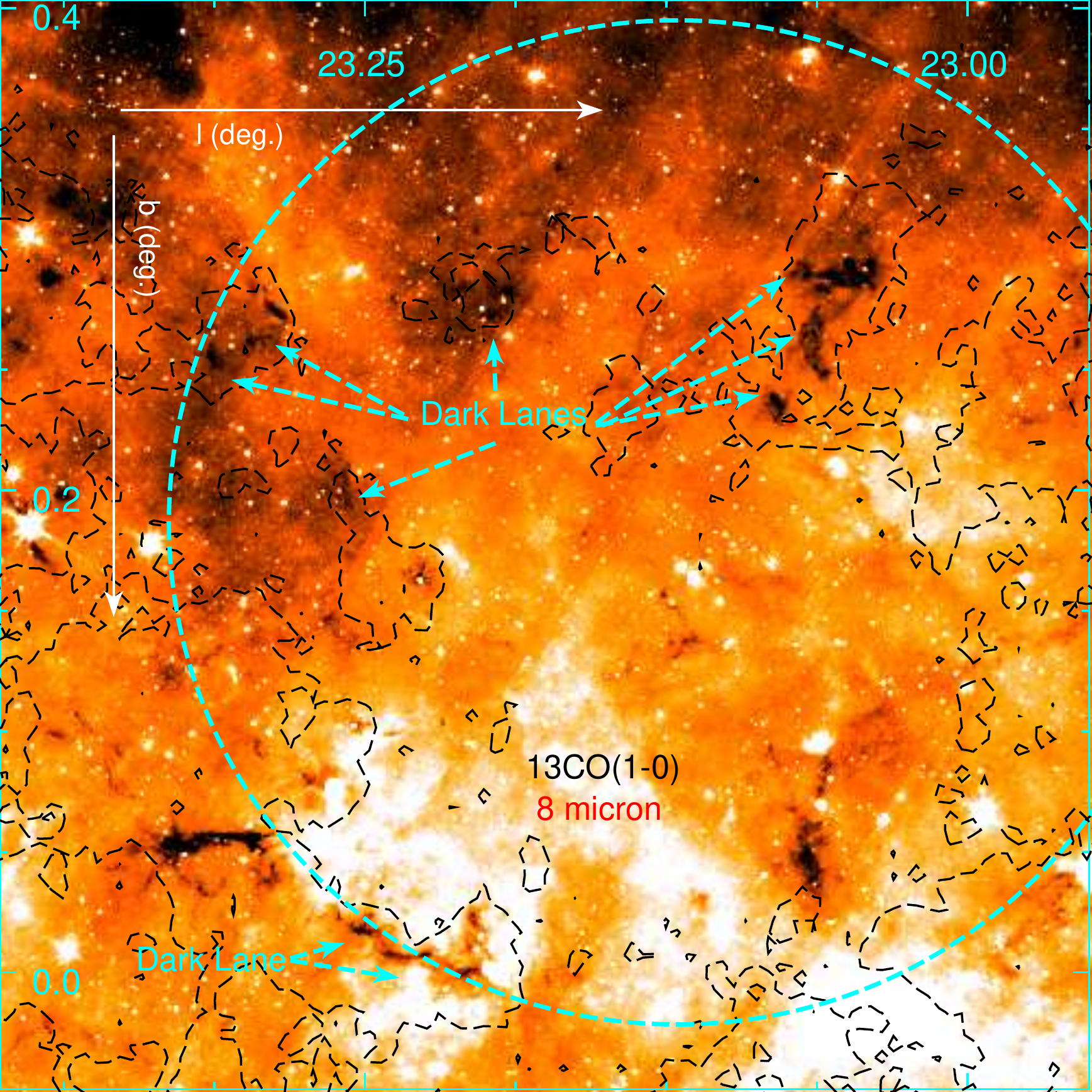}
\caption{\textbf{Top left:} Three-colour image of FUGIN $^{12}$CO(1-0) (red), $^{13}$CO(1-0) (green), and C$^{18}$O(1-0) (blue) emission, velocity-integrated between 70 and 90\,km~s$^{-1}$. HESS 3, 4, 5, 6 and 7$\sigma$ contours are overlaid. A dashed vertical line indicates the boundary between finalised (left) and preliminary (right) image processing properties. The preliminary data exhibits some striping artefacts. \textbf{Top right:} Two-colour image of $^{13}$CO(1-0) velocity-integrated between 70 and 90\,km~s$^{-1}$ (green) and MWA 170-231\,MHz radio continuum (blue). HESS 3, 4, 5, 6 and 7$\sigma$ contours are overlaid. \textbf{Bottom} Spitzer 8$\mu$m infrared emission with overlaid FUGIN $^{13}$CO(1-0) 70-90\,km~s$^{-1}$ emission 15~K\,km~s$^{-1}$ contours. {Dark lanes that are morphologically suggestive of association with the $^{13}$CO(1-0) emission is indicated.}
 \label{fig:COthreeCol}}
\end{figure*}

We did, however, find infrared-dark features that match the morphology of molecular gas traced by \hbox{FUGIN} CO. In Figure\,\ref{fig:COthreeCol}, we display 8\,$\mu$m emission with $^{13}$CO(1-0) emission contours from the 70-90\,km~s$^{-1}$ velocity slice overlaid. One $^{13}$CO clump in the north-west of \gtt\ has a notable corresponding dark-lane association, while several dark lanes lie within $^{13}$CO-traced gas at the south-eastern boundary. These associations suggest \newtext{that the gas identified as a potential association for \gtt\ is foreground to a significant number of stars within Galactic plane, hence favouring a near-side solution to the kinematic degeneracy of Galactic rotation in the inner plane. }
This is an important clue for determining the SNR distance, because we expect components of the near and far Norma-arms to be confused with the far-Scutum-Crux arm for velocities above $\sim$75\,km/s (e.g. see Figure\,\ref{fig:PVplots_lon}).

\subsection{A Search for \gtt\ in X-rays}\label{sec:Xray}

\begin{figure}
\centering
\includegraphics[width=0.48\textwidth]{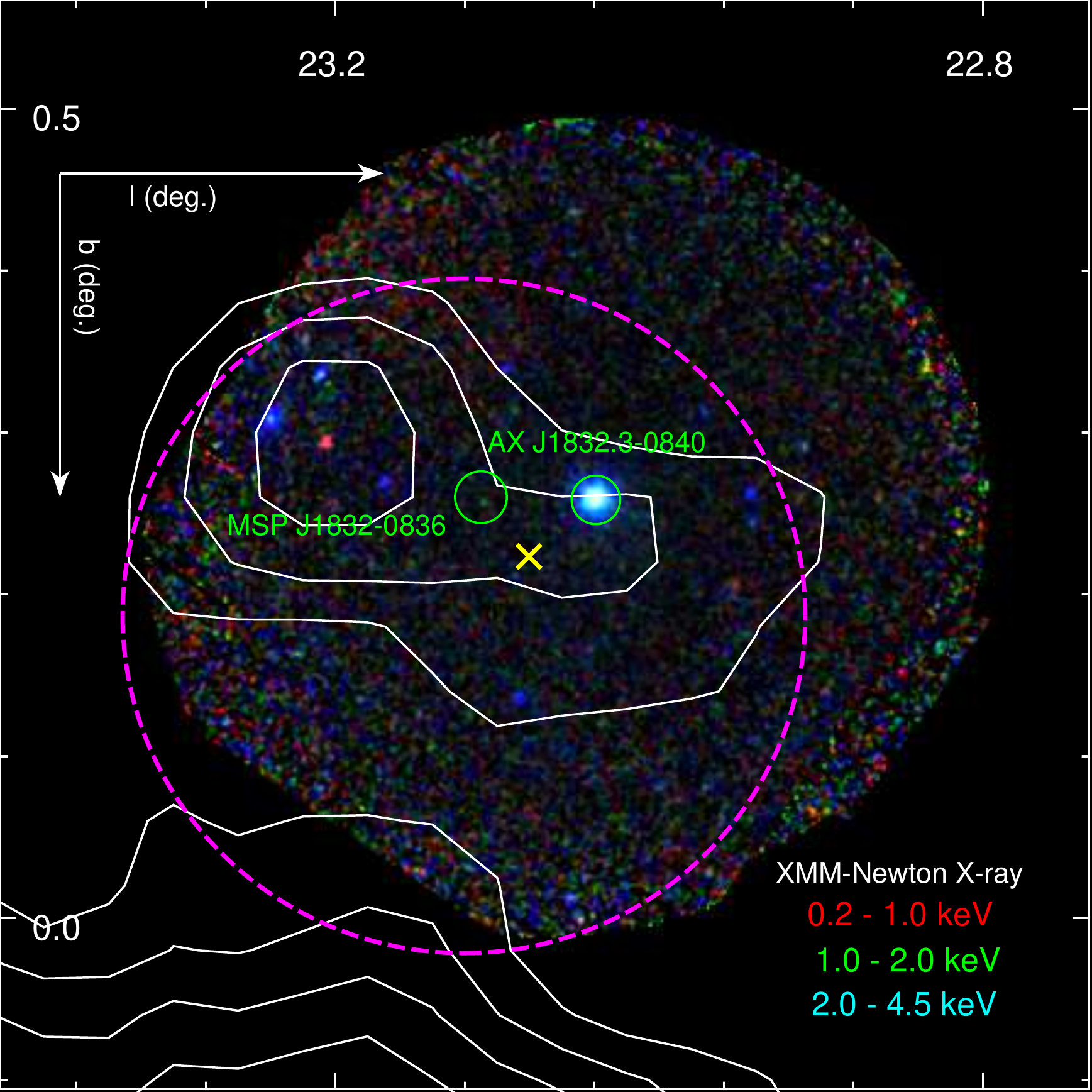}
\caption{The {\it XMM-Newton} three-colour image showing 0.2-1, 1.0-2.0 and 2.0-4.5~keV bands in red, green and blue, respectively. HESS 3, 4 and 5$\sigma$ $>$1\,TeV gamma-ray emission contours are overlaid \citep{Abdalla:2018}. The position of \gtt\ is indicated by a broken magenta circle. The positions of the cataclysmic variable star AX~J1832.3-0840 and millisecond pulsar MSP\,J1832$-$0.836 \citep{0004-637X-864-1-23} are indicated by green circles. A yellow cross marks the radio continuum source possibly associated with G023.090$+$0.220 
 \label{fig:xmm}}
\end{figure}

In an attempt to establish the \gtt\ candidacy as an SNR, we examined \textit{XMM-Newton} X-ray data. In Figure~\ref{fig:xmm}, we present X-ray images of the 0.2-1~keV, 1.0-2.0~keV and 2.0-4.5~keV bands, which represent `soft', `medium' and `hard' X-ray emission, respectively. No significant diffuse X-ray emission is seen in the images, \newtext{or in higher energy bands of $\sim$4.5-10\,keV}, and we are unable to reinforce the evidence in favour of the SNR candidate \gtt\ using this X-ray data-set. \newtext{We attribute this to a likely high level of foreground absorption (see Section\,\ref{sec:Xray}).}\footnote{We note that a large number of prominent radio SNRs are not readily detected in X-ray images \citep[e.g.][]{2010A&A...518A..35C}.}

\subsection{The Nature and Distance of \gtt}
 \label{sec:Distance}
In Section\,\ref{sec:Gas}, we examined position-velocity plots to identify two voids in molecular gas, Void\,A in the Norma arm and Void\,B in the Scutum-Crux arm. {We suggested these to be associations} for SNR candidate \gtt . Then, under the assumption that the gamma-ray source \hje\ is associated with the SNR, we suggested that gas in the velocity range neighboring Void\,A is also associated with \gtt , which would favour the Void\,A association interpretation for \gtt\ over Void\,B. Furthermore, the Void\,B interpretation is disfavoured by the possibility that one side of this structure may in fact be part of a confused far-side Galactic arm - the far-Sagittarius arm, perhaps tentatively observed in Figure\,\ref{fig:PVplots_lon} at $\sim$65\,km~s$^{-1}$. With this, and the lack of gamma-ray/gas correspondence in the Scutum-Crux arm in mind, we move forward with a Clump\,A-Void\,A \gtt\ association hypothesis.

\newtext{The proposed scenario presents a consistent picture that unifies the radio continuum remnant, the gamma-ray source and the ISM. In this interpretation, the wind of the \gtt\ progenitor star blows-out a cavity in nascent Norma-arm molecular gas before exploding within the region evacuated of diffuse gas. The supernova remnant shell, seen in radio continuum, expands to interact with clumps remaining within the cavity and eventually the cavity wall. Particle acceleration occurs and initiates gamma-ray emission, which may either correspond to escaping CR hadrons interacting with northern clouds \citep[e.g.][]{Aharonian:w28} adjacent in the line of sight, or CR hadrons interacting with clumps within the SNR shell \citep[e.g.][]{Fukui:2012}. Both of these scenarios are consistent with a hadronic gamma-ray emission mechanism for \gtt . A leptonic scenario may also be consistent with the proposed ISM association if the population of high energy electrons is boosted in regions where the SNR shock meets northern gas, as briefly suggested by \citet[][]{Sushch:2018} for the young SNR Vela\,Jr. In the Vela\,Jr case, however, TeV gamma-ray emission was accompanied by strong non-thermal X-ray emission from the same population of $>$TeV electrons - something not seen in \gtt , which may have a less-energetic population of electrons due to a larger age, as modelled below.  In either case (leptonic or hadronic), the Clump\,A-Void\,A \gtt\ association hypothesis is consistent irrespective of the gamma-ray emission mechanism. }

Assuming a Milky Way Galactic centre distance and Galactic rotation Velocity of 8\,kpc and 230\,km~s$^{-1}$, respectively \citep[see][]{Vallee:2017}, we estimate the kinematic distance of the Void\,A+Clump\,A gas (70-90\,km~s$^{-1}$) to be \distance\ \citep{Brand:1993}, where the distance range corresponds to the velocity ranges of the features in question, while the statistical error of 0.4\,kpc corresponds to 10\,km\,s$^{-1}$ (conservatively estimated as the expectation for local gas motions). Here, we are assuming the near-side solution, due to the presence of dark lanes morphologically matching the 70-90\,km~s$^{-1}$ gas \newtext{(see Section\,\ref{sec:Infrared})}. There is some inconsistency between the calculated kinematic distance of \distance , and the schematic in Figure\,\ref{fig:PVplots_lon}, which would place gas at this distance between Galactic arms. In contrast, we consider the ordering and velocity of clear, significant Galactic arms in the position-velocity plots to be a more reliable tool for assigning a Galactic arm to the 70-90\,km~s$^{-1}$ gas, in this case the Norma arm.

Under the assumption of an association between the SNR candidate \gtt\ and the CO-traced molecular gas of the Norma-arm, the \distance\ distance can also be applied to \gtt , to derive a SNR diameter of {31.5$\pm$7.9\,pc (for a SNR angular diameter of 21.7-25.0$^{\prime}$, {and including the kinematic distance uncertainty)}}. Furthermore, the wind-blown bubble nature implied by such an association would mean that the SNR originates from a core-collapse progenitor event. 

We model the feasibility of this scenario using SNR evolution modeling software by \citet{Leahy:2017}, which primarily utilises the approach outlined in \citet{Truelove:1999}. We assume a low explosion energy of 0.5$\times$10$^{51}$\,erg, consistent with a core-collapse event.\footnote{We also assume an ejecta mass of 2\,M$_{\odot}$ - a parameter that subsequent results are much less sensitive to than explosion energy and ISM density.} 

With the assumed diameter of 31.5$\pm$7.9\,pc, the SNR age could range from 2.1-5.2\,kyr if the evolution takes place in an evacuated medium of homogeneous density 0.01\,cm$^{-3}$, to an age of 8.6-54\,kyr if inside a dense 1.0\,cm$^{-3}$ medium. In the former case, strong non-thermal X-ray emission would be produced from a population of 1-10\,keV electrons within the fast $>$1000\,km\,s$^{-1}$ Sedov-Taylor-phase shocks of \gtt, while the latter case would result in a significant 10-100\,eV thermal X-ray component as radiative-phase shocks slow to velocities of 100-300\,km~s$^{-1}$ range. Such soft X-ray emission, however, \newtext{might be difficult to detect due to high foreground photoelectric absorption (see Section\,\ref{ssec:Xrayobs}).}
\newtext{
The non-detection of strong non-thermal emission in \textit{XMM-Newton} data, including above $\sim$5\,keV where photoelectric absorption is negligible, favours a middle age ($\sim$10$^4$\,yr) for \gtt . \gtt\ is thus probably not similar to the young, $\sim$10$^3$yr-age, radio-dim, gamma-ray-bright shell-type SNRs such as RX\,J1713.7$-$3946 or Vela\,Jr\footnote{We further note that the group of TeV SNR shells \citep{Abramowski:2017newshells}, which \gtt\ may be a part of, are prominent examples where SNR evolution is not fully understood.}. If \gtt\ originated from a similar type of core-collapse explosion into an evacuated cavity, the SNR is likely now at a more-advanced stage of evolution, consistent with the larger diameter ($\sim$32\,pc) at the proposed distance of the CO(1-0) void ($\sim$4.6\,kpc). Furthermore, the lack of non-thermal keV emission towards \gtt\ indicates that a substantial population of $>$1\,TeV electrons is not present in the region, thus disfavouring a leptonic scenario for gamma-ray emission, adding weight to the proposed existence of a residual population of $>$TeV CR protons.} 

No GeV gamma-ray detection is recorded towards \gtt\ in the most recent Fermi-LAT catalogue \citep{FermiLAT:2015}. The \hje\ $>$1\,TeV spectral index is 2.38$\pm$0.14, consistent with HESS TeV gamma-ray shells such as Vela\,Jr, RX\,J1713.7$-$3946, HESS\,J1731$-$347, RCW\,86, SN\,1006 and HESS\,J1534$-$571 which have TeV spectral indices of 2.3-2.5 \citep[e.g. see Tables 4 and 6 of][]{Abramowski:2017newshells}. The flux of \hje , however, is 0.9$\pm$0.2\%\,Crab\footnote{$>$1\,TeV gamma-ray fluxes are often expressed as a fraction of the TeV-bright Crab SNR, e.g. see \citet{Aharonian:2006crab}.} --  smaller than these HESS TeV shells. For comparison, if \hje\ is indeed at the distance of \distance , its $>$1\,TeV luminosity is $\sim$20-35\% that of RX\,J1713.7$-$3946. 

We caution the reader that the proposed association between the SNR candidate, gamma-ray source and ISM might simply be a line-of-sight effect in a scenario where the \hje\ gamma-ray emission originates from runaway CRs produced by SNR W41 rather than SNR candidate \gtt\ (see Section\,\ref{sec:w41}). {This scenario is not favoured by the CO distribution at any line-of-sight velocity in an isotropic CR diffusion scenario, but might still be feasible if anisotropic diffusion effects disproportionately increase the CR density directly north of W41, possibly due to an ordered magnetic-field structure \citep[e.g.][]{Nava:2013,Lau:2017}. }

Finally, we note that the pulsar PSR\,J1832$-$0827 \citep{Abramowski:2015_HESSJ1832-085}, is at a distance of $\sim$4.9\,kpc \citep[][also see \citealt{Clifton:1986,Frail:1991}]{Cordes:2002} - compatible with the ISM association suggested by our study. The characteristic age of $\sim$200\,kyr \citep[as estimated from the spin-down luminosity and braking index,][]{Hobbs:2004,Johnston:1999}, however, is not consistent with a \gtt\ association. At such an age the SNR would likely be beyond the radiative phase and merged with the ISM,\footnote{As can be observed by exploring both reasonable and unreasonable parameter spaces using the \citet{Leahy:2017} SNR evolution modelling software.} hence not detectable. The pulsar ages calculated from spin-down luminosity, however, are generally not reliable measures of real age, because energy loss-rates are not expected to be constant due to likely variations in magnetic field strength \citep[e.g. see][]{Chanmugam:1995} and cases of inconsistency between SNR age and pulsar spin-down age are well-known \citep[e.g.][]{Gotthelf:2009}. {Furthermore, the prospect of the pulsar being `spun-up' complicates the age estimate further. For these reasons,} we do not rule-out an association between \gtt\ and PSR\,J1832$-$0827.



\section{CONCLUSION}
We present an examination of the new Galactic SNR candidate \gtt. MWA radio continuum data indicate a shell-like morphology and a spectral index of $\alpha$=--0.63$\pm$0.05, while the \newtext{coincident} TeV gamma-ray source \hje\ towards the northern hemisphere of \gtt\ {confirms that the object is a likely SNR and viably} accelerating particles to super-TeV energies. We conduct an investigation into the nature of the source using archival data. Although we find no corresponding X-ray, infrared or optical emission, based on morphological considerations we propose an association with molecular gas at a kinematic distance of \distance . We propose that a dip in CO-traced molecular gas at a line-of-sight velocity of $\sim$85\,km\,s$^{-1}$ is the \gtt\ progenitor wind-blown bubble, while good correspondence between \hje\ and gas at a neighbouring velocity is consistent with a gamma-ray production mechanism that involves gas, perhaps p-p interactions by cosmic rays, towards the north of the remnant. 

\section{acknowledgements}
{This work is supported by the Australian Research Council grants FT170100243 and FT160100028 and makes use of the Murchison Radio-astronomy Observatory, operated by CSIRO. We acknowledge the Wajarri Yamatji people as the traditional owners of the Observatory site. Support for the operation of the MWA is provided by the Australian Government (NCRIS), under a contract to Curtin University administered by Astronomy Australia Limited. We acknowledge the Pawsey Supercomputing Centre which is supported by the Western Australian and Australian Governments. This work was compiled in the very useful free online LaTeX editor Overleaf. \newtext{Finally, we thank the anonymous referee, who provided very constructive and helpful comments that improved the quality of our manuscript.} }

%




\newtext{\software{\textit{Miriad} \citep{Sault:1995},
\textit{ds9} visualization application \citep{Joye:2003}, The \textit{Karma} package \citep{Gooch:1995,Gooch:1996}.}}

\newpage

\bibliographystyle{aasjournal}
\bibliography{Refs_SNRg23}



\end{document}